\documentclass[twocolumn]{aastex63}
\shorttitle{The Dwarf Galaxy Population at $z\sim 0.7$}
\shortauthors{Pharo et al.}
\graphicspath{{./}{figures/}}
\begin{document}

\title{The Dwarf Galaxy Population at $z\sim 0.7$: A Catalog of Emission Lines and Redshifts from Deep Keck Observations}

\correspondingauthor{John Pharo}
\email{jp7tp@missouri.edu}

\author{John Pharo}
\affiliation{Department of Physics and Astronomy, University of Missouri, Columbia, MO 65211, USA}

\author{Yicheng Guo}
\affiliation{Department of Physics and Astronomy, University of Missouri, Columbia, MO 65211, USA}

\author{Guillermo Barro Calvo}
\affiliation{University of the Pacific, Stockton, CA, USA}

\author{Timothy Carleton}
\affiliation{School of Earth and Space Exploration, Arizona State University, Phoenix, AZ, USA}

\author{S. M. Faber}
\affiliation{UCO/Lick Observatory, Department of Astronomy and Astrophysics, University of California, Santa Cruz, CA, USA}
\affiliation{University of California, Santa Cruz, CA, USA}

\author{Puragra Guhathakurta}
\affiliation{UCO/Lick Observatory, Department of Astronomy and Astrophysics, University of California, Santa Cruz, CA, USA}
\affiliation{University of California, Santa Cruz, CA, USA}

\author{Susan A. Kassin}
\affiliation{Space Telescope Science Institute, 3700 San Martin Drive, Baltimore, MD 21218, USA}
\affiliation{Department of Physics and Astronomy, Johns Hopkins University, 3400 N. Charles Street, Baltimore, MD 21218, USA}

\author{David C. Koo}
\affiliation{UCO/Lick Observatory, Department of Astronomy and Astrophysics, University of California, Santa Cruz, CA, USA}
\affiliation{University of California, Santa Cruz, CA, USA}

\author{Jack Lonergan}
\affiliation{Department of Physics and Astronomy, University of Southern California, Los Angeles, CA, 90007, USA}

\author{Teja Teppala}
\affiliation{Department of Physics and Astronomy, University of Missouri, Columbia, MO 65211, USA}

\author{Weichen Wang}
\affiliation{Department of Physics and Astronomy, Johns Hopkins University, 3400 N. Charles Street, Baltimore, MD 21218, USA}

\author{Hassen M. Yesuf}
\affiliation{University of California, Santa Cruz, CA, USA}
%\affiliation{University of California at Santa Cruz, Santa Cruz, CA, USA}

\author{Fuyan Bian}
\affiliation{European Southern Observatory, Chile}

\author {Romeel Dav\'e}
\affiliation{University of Edinburgh, Edinburgh, United Kingdom}

\author{John C. Forbes}
\affiliation{Center for Computational Astrophysics at the Flatiron Institute, New York, NY, USA}

\author{Dusan Keres}
\affiliation{University of California, San Diego, La Jolla, CA, USA}

\author{Pablo Perez-Gonzalez}
\affiliation{Centro de Astrobiolog\'ia (CAB, INTA-CSIC), Carretera de Ajalvir km 4, E-28850 Torrej\'on de Ardoz, Madrid, Spain}

\author{Alec Martin}
\affiliation{Department of Physics and Astronomy, University of Missouri, Columbia, MO 65211, USA}

\author{A. J. Puleo}
\affiliation{Department of Physics and Astronomy, University of Missouri, Columbia, MO 65211, USA}

\author{Lauryn Williams}
\affiliation{Department of Physics and Astronomy, University of Missouri, Columbia, MO 65211, USA}

\author{Benjamin Winningham}
\affiliation{Department of Physics and Astronomy, University of Missouri, Columbia, MO 65211, USA}



\begin{abstract}

We present a catalog of spectroscopically measured redshifts over $0 < z < 2$ and emission line fluxes for 1440 galaxies. The majority ($\sim$65\%) of the galaxies come from the HALO7D survey, with the remainder from the DEEPwinds program. This catalog includes redshifts for 646 dwarf galaxies with $\log(M_{\star}/M_{\odot}) < 9.5$. 810 catalog galaxies did not have previously published spectroscopic redshifts, including 454 dwarf galaxies. HALO7D used the DEIMOS spectrograph on the Keck II telescope to take very deep (up to 32 hours exposure, with a median of $\sim$7 hours) optical spectroscopy in the COSMOS, EGS, GOODS-North, and GOODS-South CANDELS fields, and in some areas outside CANDELS. We compare our redshift results to existing spectroscopic and photometric redshifts in these fields, finding only a 1\% rate of discrepancy with other spectroscopic redshifts. We measure a small increase in median photometric redshift error (from 1.0\% to 1.3\%) and catastrophic outlier rate (from 3.5\% to 8\%) with decreasing stellar mass. We obtained successful redshift fits for 75\% of massive galaxies, and demonstrate a similar 70-75\% successful redshift measurement rate in $8.5 < \log(M_{\star}/M_{\odot}) < 9.5$ galaxies, suggesting similar survey sensitivity in this low-mass range. We describe the redshift, mass, and color-magnitude distributions of the catalog galaxies, finding HALO7D galaxies representative of CANDELS galaxies up to \textit{i}-band magnitudes of 25. The catalogs presented will enable studies of star formation (SF), the mass-metallicity relation, SF-morphology relations, and other properties of the $z\sim0.7$ dwarf galaxy population.

\end{abstract}



\section{Introduction} \label{sec:intro}

\begin{figure*}[t!]
    \centering
    \includegraphics[width=0.75\textwidth]{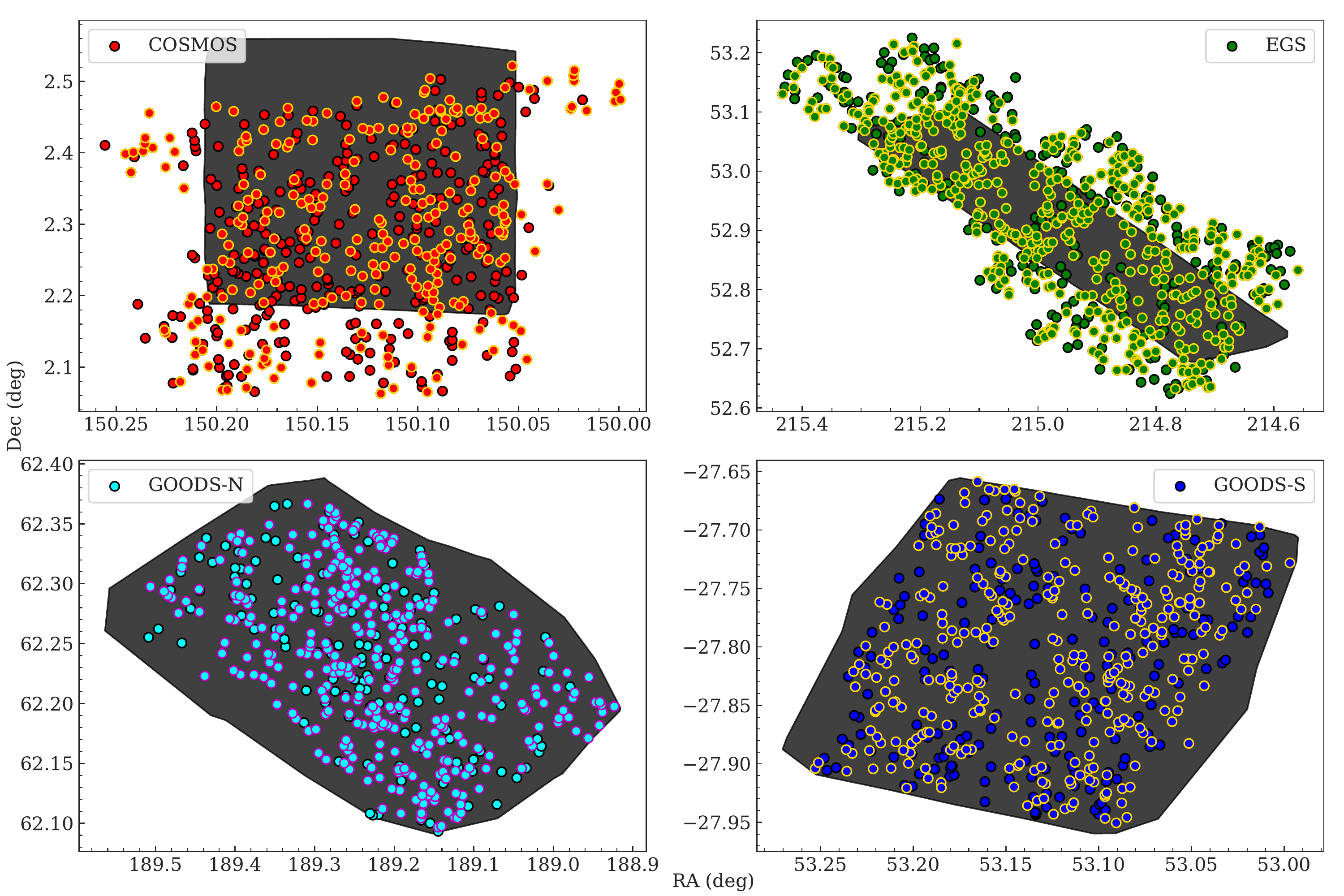}
    \caption{The 4 fields targeted in the various HALO7D observations, clockwise from top left: COSMOS, EGS, GOODS-South, GOODS-North. Black shaded regions indicate coverage from the CANDELS catalogs in these fields \citep{dahlen13,guo13,santini15,kocevski17,barro19} with measured photometry and redshifts. The colored points indicate the locations of HALO7D target galaxies. We obtained successful redshift fits for points with colored outlines (gold or magenta), and were unable to measure redshifts in HALO7D spectra for the points with black outlines. HALO7D targets in the COSMOS and EGS fields included a number of galaxies outside the regular CANDELS coverage, as can be seen where the HALO7D targets do not overlap the black CANDELS region. We obtained photometry for these galaxies from COSMOS/UltraVista \citep{muzzin13} and EGS/IRAC \citep{barro11a,barro11b} catalogs.}
    \label{fig:fields}
\end{figure*}

Low-mass or ``dwarf'' galaxies, those with stellar masses of $10^{9.5} M_{\odot}$ or less, are a critical galaxy population for understanding galaxy evolution, and may serve as effective probes of the mechanisms behind star formation and metallicity enrichment in particular. On the tight relation between the stellar mass ($M_{\star}$) and the star formation rate, sometimes called the star-forming main sequence, low-mass galaxies exhibit higher star formation rates (SFR) per stellar mass, or sSFR, than their more massive counterparts \citep{noeske07}. In addition to higher levels of star formation, in the local Universe, several previous studies have found evidence that bursty star formation, wherein star formation is rapidly triggered and quenched on a timescale of tens of Myr, is more prominent among low-mass galaxies \citep[e.g.,][]{searle73,bell01,lee09,meurer09,weisz2012}. Theoretical models relating stellar mass and gas-phase metallicity with supernova-driven galactic winds also predict low-mass galaxies will exhibit more scatter in the Mass-Metallicity and Fundamental Metallicity Relations \citep{henry13,lu15,guo16a}, empirical correlations observed between stellar mass and star formation to the galaxy's gas-phase metallicity. These relations and their intrinsic scatter are compared with theoretical predictions, such as the outflow models described above, to probe the gas flows or other physical mechanisms regulating star formation \citep{dave12, lilly13, forbes14b}. This regulator model could relate gas flows to SFR variability and consequently to star formation burstiness \citep[e.g., ][]{tacchella2020}, so a more detailed study of the particularly bursty and high-scatter dwarf galaxy population is key to testing the vailidity of such models. Furthermore, this expected scatter introduces potential uncertainty in photometric redshift measurements of dwarf galaxies, as the galaxy templates used for redshift fitting are more likely to be representative of more massive galaxies that are more commonly measured spectroscopically.

Optical and infrared spectroscopy of galaxies have been used to great success in studying these relations in galaxies, including in the local universe; at $0.5 < z < 2$ as in e.g., DEEP2 \citep{newman13} and DEEP3 \citep{cooper12} with Keck/DEIMOS, LEGA-C with VLT/VIMOS \citep{vanderwel16}, and with VLT/MUSE \citep{inami17, carton18, urrutia2019} and KMOS \citep{gillman21}; and at $z > 2$ in e.g., MOSDEF \citep{kriek2015} and KBSS-MOSFIRE \citep{rudie2012}. Spectroscopy simultaneously provides precise identification of a galaxy's redshift, a necessary requirement for the determination of the stellar mass via SED fitting, and the means to measure star formation \citep{kennicutt98}, gas-phase metallicity \citep[e.g., ][]{kewley01}, and other properties derived from the fluxes of nebular emission lines \citep{kewley19} or features of the stellar continuum. Consequently, observations capturing the rest-frame optical spectrum of a galaxy can yield a wealth of information on that galaxy's properties and star-forming characteristics, provided enough signal and resolution to detect the line emission. 

Deep spectroscopy of low-mass galaxies at higher redshift will enable the study of these star formation properties among dwarf galaxies closer to the $z \sim 2$ epoch of peak SFR \citep{madau14}, as well as among the sample of galaxies which are likely progenitors of Milky-Way-like galaxies in the local Universe \citep{vandokkum13,papovich15}. Rest-optical spectroscopy of dwarf galaxies will also provide a key test of the accuracy of photometric redshift measurements of dwarfs, with implications for both the study of the above relations as well as large-scale structure studies. However, the low-mass galaxy population has not been thoroughly studied at $ z > 0.5$ due to the difficulty in obtaining sufficiently deep spectroscopy to measure star formation in such faint targets. For example, \citet{guo16b} measured star formation burstiness with Keck/DEIMOS spectroscopy at $0.5 < z < 1$ for 164 galaxies with $ 8.5 < \log(M_{\star}/M_{\odot}) < 10.5$, but only 17 galaxies in the sample had $\log(M_{\star}/M_{\odot}) < 9$. Grism spectroscopy can be a means to achieve the necessary depth; \citet{pharo20} measured emission lines for $\sim50$ low-mass galaxies at $0.3 < z < 2$ with Hubble Space Telescope WFC3-G102 near-infrared grism spectroscopy with 40-orbit depth \citep[(FIGS)][]{pirzkal2017}, but limitations of survey area, wavelength coverage, and spectral resolution make the sample sizes obtainable in previous studies below what is needed for meaningful studies of the SFR properties of low-mass galaxies.

With deep optical spectroscopy from the HALO7D program \citep{cunningham2019a}, we are able to achieve both sufficiently high signal-to-noise and sufficiently large survey size needed to securely identify and measure a population of hundreds of low-mass galaxies with redshifts $0 < z < 1.6$. HALO7D consists of Keck/DEIMOS optical spectroscopy with up to 32 hours of exposure in four of the CANDELS fields \citep{grogin11, koekemoer11}. 

%-understanding the nature of star formation in dwarf galaxies is a critical aspect of galaxy evolution

%-this has been studied in the Local universe, but it is difficult to do so at higher redshift

%-studying the dwarf population at higher redshift is necessary to test our understanding of SF in a distinct epoch from the local Universe, and a time when SF density was higher

%-spectroscopy of emission lines is a reliable and effective way to study these properties

%-however, it is typically difficult to do spectroscopy of faint dwarf galaxies

%-with HALO7D, we bridge this gap, and present a catalog of reliable galaxy redshifts and emission lines derived from deep Keck/DEIMOS optical spectra, with which we are able to characterize $\sim$350 dwarf galaxies with $0.4 < z < 1$.

\begin{figure*}[t!]
    \centering
    %\begin{tabular}{c}
    %    \includegraphics[width=0.95\textwidth]{Figures/exspec_low_mass.png} \\
    %    \includegraphics[width=0.95\textwidth]{Figures/exspec_high_mass_sf.png} \\
    %    \includegraphics[width=0.95\textwidth]{Figures/exspec_high_mass_q.png}
    %\end{tabular}
    \includegraphics[width=\textwidth]{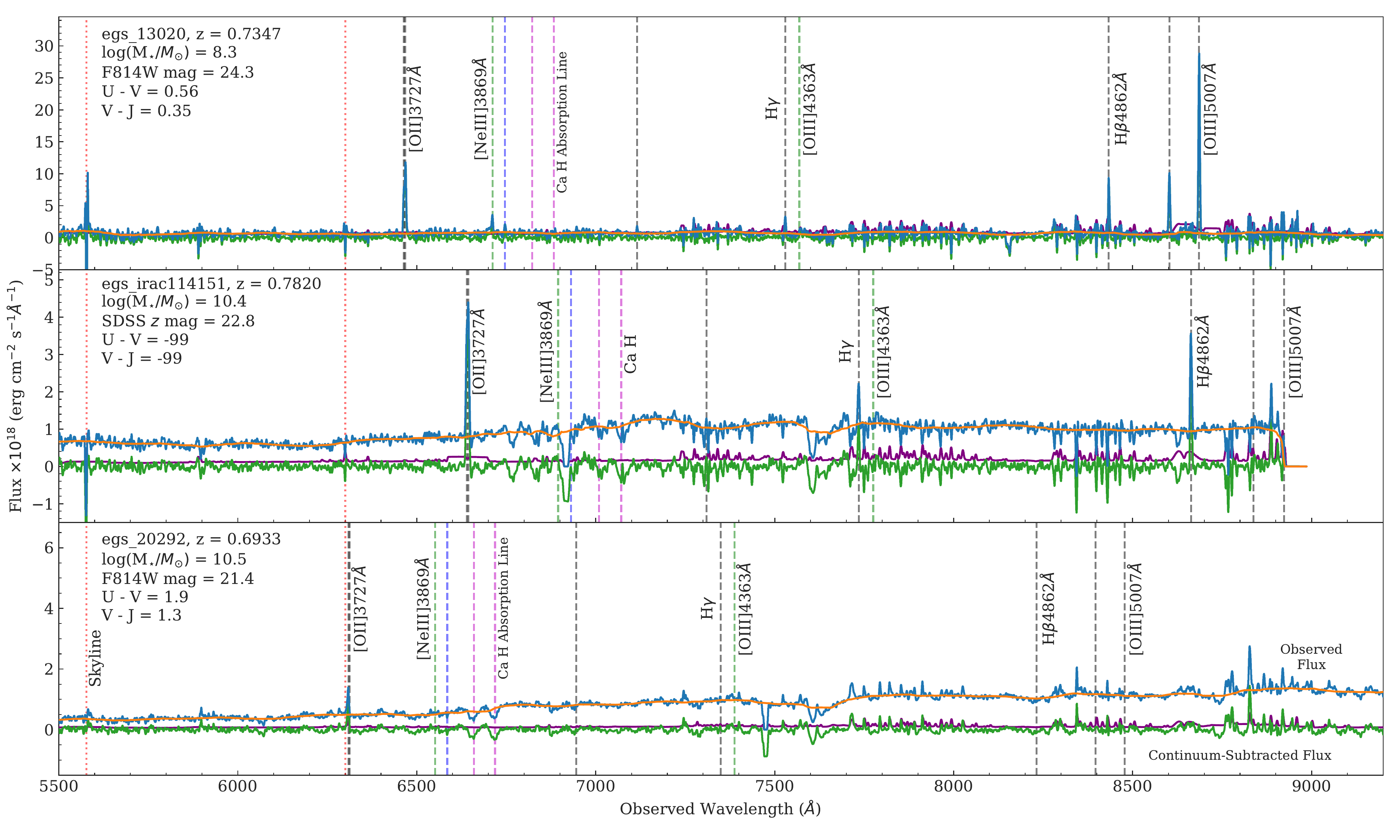}
    \caption{\textit{Top: }An example 1D spectrum of a low-mass galaxy ($\log(M_{\star}/M_{\odot}) \approx 8.3)$) in the EGS field. The blue shows the coadded spectrum of all 1D extractions, and the green shows the continuum-subtracted spectrum, though the observed difference between the two is small, as the estimated continuum (orange line) is quite small for such a low-mass galaxy. The redshift routine searched the continuum-subtracted spectrum for significant residual flux peaks in order to identify emission lines. The error spectrum is shown in purple, generally only visible when coinciding with stronger skylines, which are more common at the redder wavelengths. The vertical dotted red lines show common strong skylines that were excluded from the emission line detections. Vertical dashed black lines show common strong emission lines (e.g., [O\textsc{iii}]4959,5007; H$\beta$; [O\textsc{ii}]3727,3729). Green vertical dashed lines show the locations of weak emission lines that may serve as metallicity indicators ([NeIII] and [O\textsc{iii}]4363), and dashed blue lines indicate typically weak Helium lines. Magenta dashed lines indicate the locations of the Ca H and K absorption lines, typically only seen in higher-mass galaxies with stronger continua and older stellar populations. \textit{Middle: }An example spectrum for a high-mass galaxy ($\log(M_{\star}/M_{\odot}) \approx 10.4)$) in EGS that nonetheless exhibits some indicators of star formation, with detectable [O\textsc{ii}]3727,3729 emission and H$\beta$ and $H\gamma$ Balmer lines. This galaxy lies outside primary CANDELS photometric coverage, and so lacks some photometric and color information. \textit{Bottom: }Another massive galaxy ($\log(M_{\star}/M_{\odot}) \approx 10.5)$) in EGS that is quiescent, with no visible Balmer emission. This galaxy's redshift is determined via the Ca H and K absorption lines.}
    \label{fig:exspec}
\end{figure*}

In this paper, we present a redshift catalog of galaxies measured with Keck/DEIMOS optical spectroscopy in the HALO7D survey and other archival Keck programs. In \S2, we describe the observations and target fields, DEIMOS instrument capabilities, and spectral extraction procedure. In \S3, we describe the inspection and coaddition of the spectra, and the method by which we identified and fit the emission lines and redshifts of the galaxies. In \S4, we investigate the accuracy and other properties of the redshift catalog, and in \S5, we analyze the redshift, stellar mass, star formation, morphology, and color-magnitude properties of the dwarf galaxy population identified in the survey. In \S6, we summarize the catalogs and our analyses, as well as listing some of the subsequent studies that will make use of them, including studies of star formation, gas-phase metallicity, and other properties of the $z\sim0.7$ dwarf galaxy population, as well as investigations of star formation burstiness, the mass-metallicity relation, and the SF-morphology relations for dwarf galaxies.

In this paper, we adopt a flat $\Lambda$CDM cosmology with $ \Omega_m = 0.3$, $\Omega_{\Lambda} = 0.7$, and the Hubble constant $H_0 = 70$ km s$^{-1}$ Mpc$^{-1}$. We use the AB magnitude scale \citep{oke83}.

\section{Data and Observations} \label{sec:samp}

\begin{figure*}
\begin{tabular}{ccc}
\centering
    \includegraphics[width=0.3\textwidth]{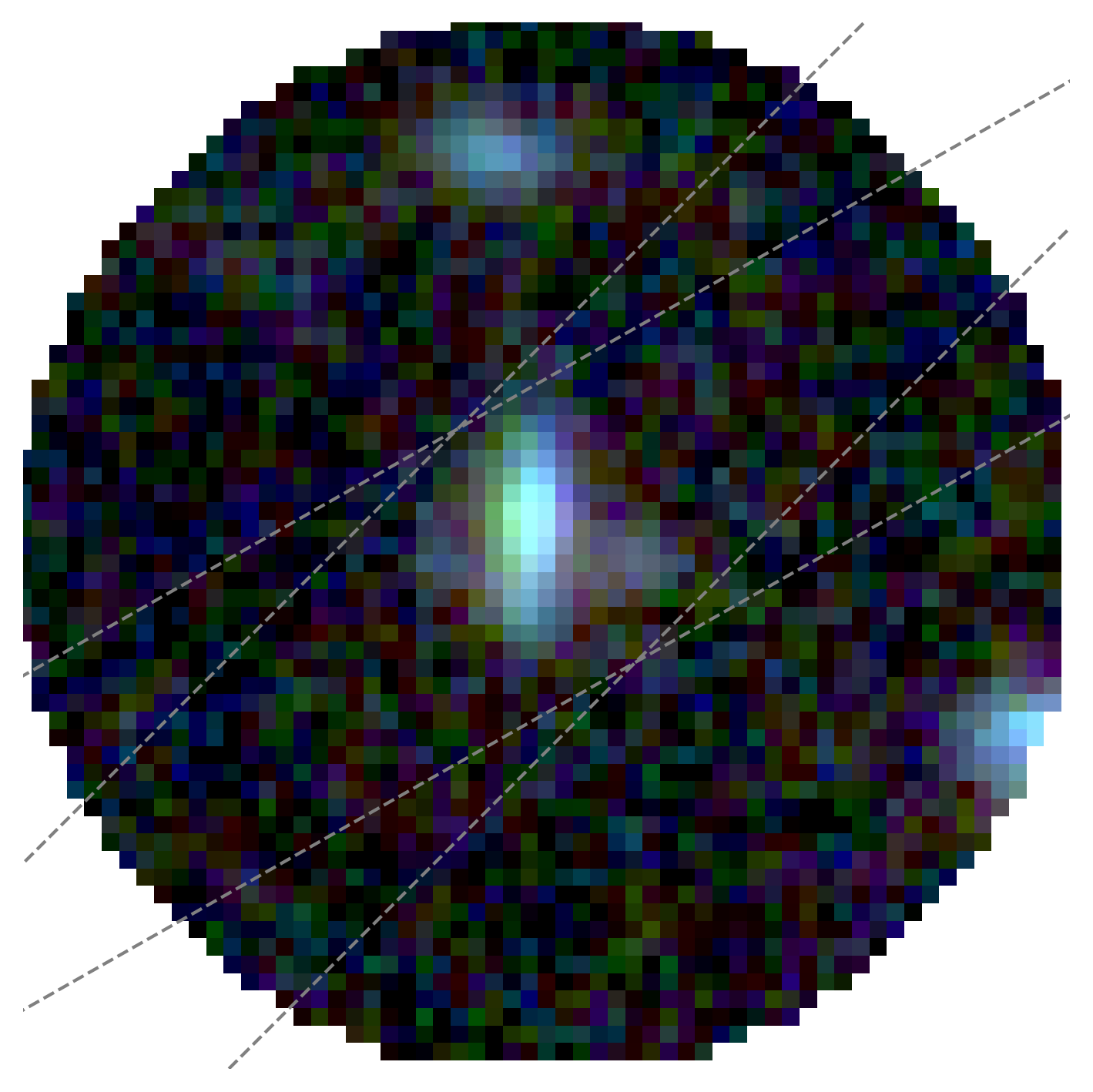} & \includegraphics[width=0.3\textwidth]{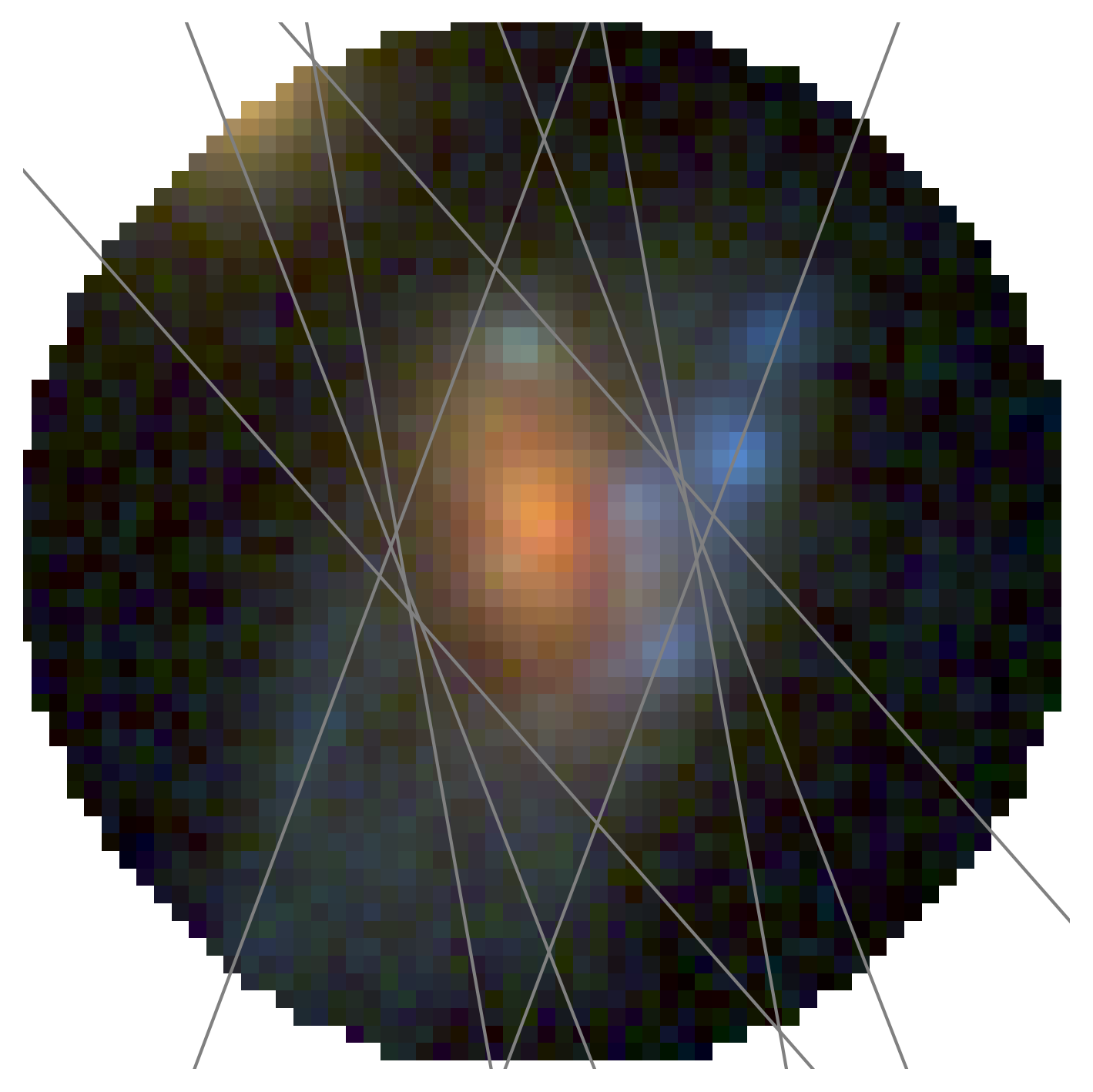} & \includegraphics[width=0.3\textwidth]{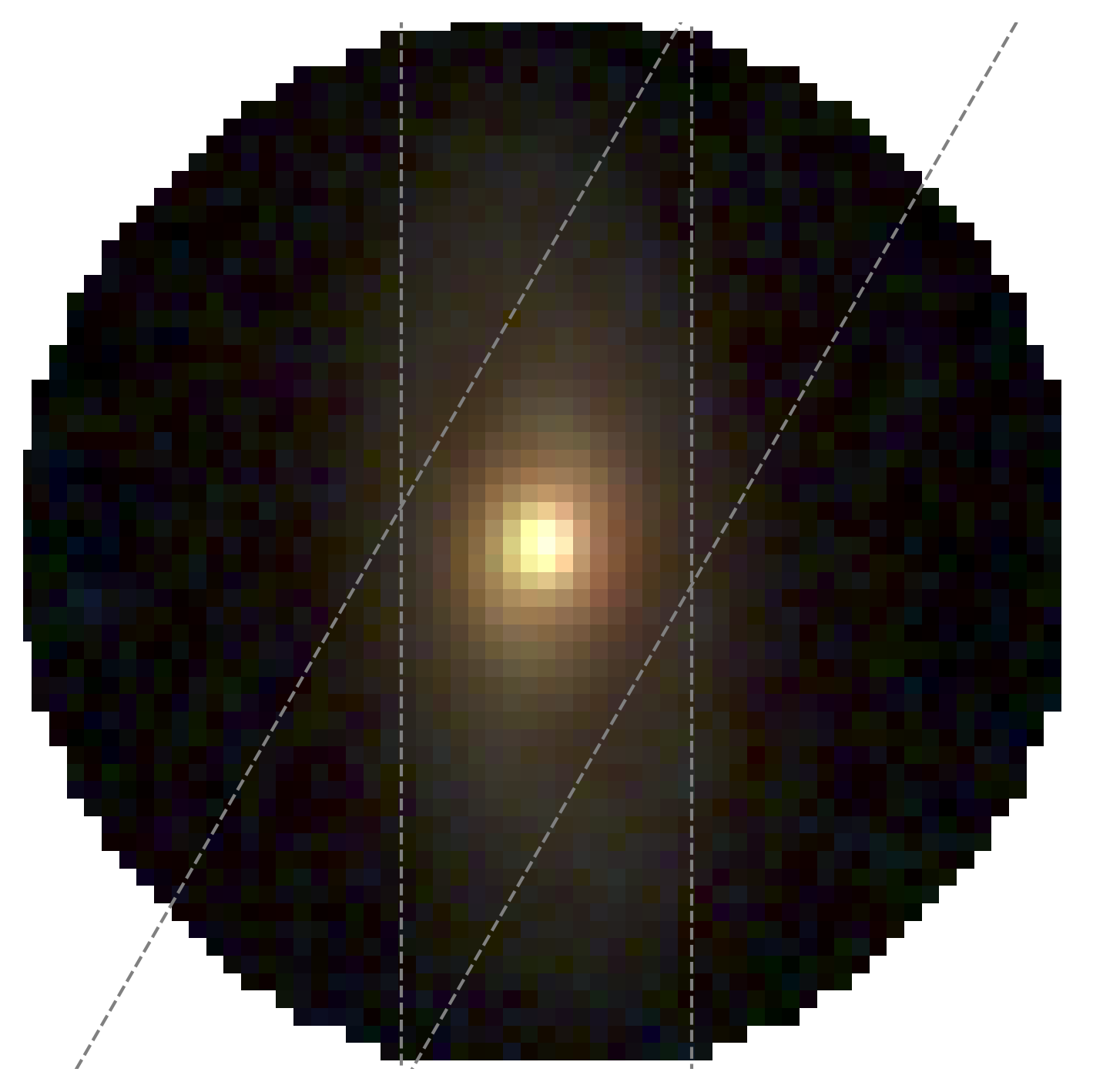}
\end{tabular}
\caption{RGB images of example HALO7D galaxies matching the three categories of target described in Figure \ref{fig:exspec}. Photometric bands used are F606W (blue), F814W/F850LP (green), and F160W (red), taken from CANDELS \citep{guo13, barro19} and 3D-HST \citep{skelton2014}. Each image has a diameter of 60 HST pixels, for a diameter of 3.6''. Dashed lines indicate the placement of spectroscopic slits.}
\label{fig:rgb}
\end{figure*}

\subsection{Data and Observational Programs} \label{sec:dp}
The data used in this paper are comprised of deep optical spectra of $\sim$2400 galaxies observed with Keck/DEIMOS. The majority of the spectra come from the HALO7D program \citep[PI: Guhathakurta;][]{cunningham2019a}, a program primarily designed to observe faint Milky Way halo stars in the COSMOS, EGS, and GOODS-North CANDELS fields \citep{grogin11, koekemoer11}. Unused space in the DEIMOS slit masks was filled out with galaxies, including a sample of 558 low-mass galaxies at $0 < z < 1.0$. Samples of high-mass galaxies were also targeted for studies of strong winds in star-forming galaxies and stellar populations in quiescent galaxies. The primary focus of this work is on the identification and properties of the low-mass sample (future papers will investigate properties such as their gas-phase metallicity and star formation), but others have used HALO7D data to explore the massive galaxies \citep{tacchella2021,wang2021}. The exposure time achieved for the low mass galaxies is 8-32 hours.

Additional programs expand the sample to include GOODS-South, including DEEPwinds (PI: S. Faber), an 8-hour survey yielding $\sim130$ low-mass ($10^8 M_{\odot} < M_{\star} < 10^9 M_{\odot}$) galaxies with F160W AB mag $ < 26.5$; and HALO7D-GOODSS (PI: Kirby), $\sim100$. The total observations comprise a sample of 2444 target galaxies, including 1255 low-mass galaxies across four CANDELS fields, as well as 1189 more massive galaxies.

%N168D (PI: Livermore), $\sim70$ galaxies;

Sky maps of the fields and HALO7D target galaxies are shown in Figure \ref{fig:fields}. Most HALO7D targets overlapped with existing CANDELS photometric coverage, so we were able to supplement our spectroscopic observations with photometry and derived properties such as photometric redshifts and stellar masses from CANDELS catalogs \citep{dahlen13,guo13,santini15,kocevski17,barro19}. In the GOODS-N and GOODS-S fields, nearly all HALO7D galaxies had a CANDELS counterpart, but in the COSMOS and EGS fields, a substantial number of galaxies were observed outside the CANDELS coverage. For these galaxies, we obtained multi-band photometry and other derived properties by crossmatching our spectroscopic catalog to the publicly available COSMOS/Ultravista catalog \citep{muzzin13} and the EGS/IRAC-selected catalog from \citet{barro11a,barro11b}. Example spectra of HALO7D galaxies are shown in Figure \ref{fig:exspec}, and RGB images are shown in Figure \ref{fig:rgb}.

%To account for this, we used photometry and derived properties from UltraVista \citep{muzzin13} and IRAC \citep{barro11a,barro11b} to supplement our spectroscopy.

\subsection{DEIMOS Spectra Properties and Extraction}

All spectra used in this project were obtained by the DEep Imaging Multi-Object Spectrograph (DEIMOS) instrument at the Keck Observatory \citep{faber03}. The Keck/DEIMOS spectrograph has an overall effective wavelength coverage of roughly $4100 < \lambda < 11000$ \AA. For the HALO7D observations, DEIMOS was configured with the 600 line mm$^{-1}$ grating centered at 7200\AA\, giving a wavelength dispersion of 0.65 \AA/pix and a usable wavelength range limited to $5000 < \lambda < 9500$ \AA\ \citep{cunningham2019a}.

The observations were reduced using the automated DEEP2/DEIMOS \textit{spec2d} pipeline developed by the DEEP2 team \citep{newman13}, described fully in \citet{yesuf17} and \citet{cunningham2019a}. Calibrations were done using a quartz lamp for flat fielding and both red NeKrArXe lamps and blue CdHgZn lamps for wavelength calibration. This yielded extracted 1D spectra for each exposure, and produced images of the reduced 2D spectra and extraction windows for the purposes of visual inspection of the data.

\section{Emission Line and Redshift Measurements} \label{sec:measure}

\subsection{Visual Inspection, Coaddition, and Flux Calibration} \label{subsec:vis}

With the reduced 2D and extracted 1D spectra for each exposure, we then began to systematically and visually examine each spectrum for observational defects, such as bad pixels or poorly subtracted skylines, which might affect the detection and measurement of emission lines. Any 2D spectrum with such an issue that might alter the line flux in the 1D extraction was flagged for removal from the final coadded spectrum of the galaxy. Some galaxies with few overall exposures had no exposures without major issues, and were removed from the catalog entirely, resulting in a sample of 2290 galaxies with at least one usable exposure.

%This visual inspection also provided the opportunity to identify emission line candidates among the overall sample of galaxies. The inspection tool included a preliminary coadd of all the observed spectra of a galaxy. A simultaneous fit to a list of N emission lines provided an initial estimate of the redshift and present emission lines, which were also recorded during the inspection.

After the completion of the visual inspections, the subset of exposures without major issues were coadded for each galaxy. Outlier pixels from individual spectra to be coadded were removed using a $2.5\sigma$ clipping from the median fluxes, excluding noisy deviations from the final spectrum. Then the spectra were combined with an inverse-variance-weighted average to produce the final coadded spectrum.

The final spectra could then be calibrated to the proper physical flux of the galaxies. For galaxies with an existing SED fit at the spectroscopic redshift, the best-fitting SED template flux is sampled at the wavelength of each pixel in the DEIMOS spectrum, and the ratio of the template flux to the observed flux at each wavelength is recorded. The median template-to-spectrum flux ratio is then used to scale the spectrum. For galaxies without an existing SED fit or for which a visual comparison of the template and the spectrum show a substantial mismatch, the spectrum was scaled directly to a linear interpolation of the broadband photometry. This primarily involved those galaxies outside the CANDELS coverage in the COSMOS and EGS fields (see Figure \ref{fig:fields}), which sometimes lacked fits in existing catalogs.

\subsection{Redshift Fitting} \label{subsec:z}

\begin{deluxetable}{ccc}
\label{tab:lines}
    \centering
    \tablecaption{Spectral Features for Redshift Fitting}
    \tablecolumns{3}
    \tablehead{
    \colhead{Line} &
    \colhead{Wavelength (\AA)} &
    \colhead{\% Detected}}
    \startdata
        Mg\textsc{ii} & 2796,2803 & $<$1 \\
        \text{[}O\textsc{ii}\text{]} & 3727,3729 & 47 \\
        Ca II K, H & 3933, 3968 & 19 \\
        H$\delta$ & 4102 & 5 \\
        H$\gamma$ & 4340 & 14 \\
        H$\beta$ & 4862 & 25 \\
        \text{[}O\textsc{iii}\text{]} & 4959,5007 & 30 \\
        H$\alpha$ & 6564 & 3 \\
        \text{[}S\textsc{ii}\text{]} & 6718,6733 & $<$1 \\
    \enddata
    
    %\tablenotetext{a}{A typically weaker or rarer feature, not used for redshift confirmation without a stronger feature}
    %\tablenotetext{b}{These lines may trace outflowing winds and thus may be offset from the systemic redshift by a few hundred km/s.}
    %\tablenotetext{c}{Absorption line}
    
\end{deluxetable}

To obtain redshift measurements from the coadded 1D galaxy spectra, we developed a routine to detect strong emission lines in a $0 < z < 2$ redshift window. This window was selected to encapsulate the region where strong line emitters were likely to be found. The [O\textsc{ii}]3727,3729 doublet moves into the IR and out of DEIMOS coverage beyond $z = 2$, and even the rarer Mg\textsc{ii}2796,2803 lines move out of detectable range by $z=2.2$. Any lines detectable at higher redshift come from the rest-UV spectrum, such as Ly$\alpha$1216 or [C\textsc{iii}]1909, are rare at this survey area and were found upon visual inspection to generate more false positive detections than plausible high-redshift candidates when included in the automated search routine, so the redshift range was capped at $z=2$. 

%First, the continuum flux was estimated and removed. At the line center wavelength predicted by the redshift, a region 20 \AA\ wide is selected to be the line region. Then the regions 50 \AA\ to either side of the line region are selected to estimate the local continuum level. A $3\sigma$ clip is applied to the fluxes to remove errant skylines or other nearby emission lines. The median of what remains is taken to be the continuum flux, and is then subtracted from both the continuum and line regions. This median subtraction is thus able to account for varying levels of continuum detection, as the median will simply scale toward 0 for low-mass galaxies with low continua (see Figure \ref{fig:exspec}), while scaling up to the detected continua of high-mass galaxies. This approach does, however, produce a very smooth continuum estimate that is not well-suited to the detection of continuum features. The standard deviation in whatever flux remains from the subtraction in the continuum region is then used to estimate the flux error per pixel.

To perform the redshift fit, first the routine estimated the continuum flux at each pixel. Because of the substantial population of faint, low-mass galaxies without well-defined stellar continua, we avoided fitting a continuum or specific continuum features. Instead, the code estimated the local continuum based on the median flux in a 50 \AA\ window to either side of the pixel, and subtracted this continuum flux. This median subtraction is thus able to account for varying levels of continuum detection, as the median will simply scale toward 0 for low-mass galaxies with low continua (see Figure \ref{fig:exspec}), while scaling up to the detected continua of high-mass galaxies. This approach does, however, produce a very smooth continuum estimate that is not well-suited to the detection of continuum features beyond strong, narrow absorption lines. The standard deviation of continuum-subtracted fluxes in this surrounding region is combined with the intrinsic flux error of the pixel (largely influenced by the skyline model) to get a flux error estimate, resulting in a continuum-subtracted residual spectrum and an error spectrum. This method of continuum-subtracted peak detection has demonstrated success in detecting emission lines from faint and low-mass sources in other deep surveys \citep[e.g.][]{yang17,pharo19,pharo20}. See Figure \ref{fig:exspec} for some example spectra showing the continuum subtraction.

Next, the continuum-subtracted spectrum was fit to a redshifted grid of emission and absorption line filters. A list of prominent spectral lines is described in Table \ref{tab:lines}: the H$\alpha$, H$\beta$, H$\gamma$, and H$\delta$ Balmer series lines; the [O \textsc{iii}] and [O \textsc{ii}] ionized oxygen lines; and the Ca H and K absorption lines. H$\alpha$ and the [S\textsc{ii}] doublet are strong enough to include in the fitting and catalog, but are rare at the redshift distribution of this sample. As part of the HALO7D target selection included AGN candidates, we include MgII emission as well, though there are few such detections. Furthermore, these lines may trace outflowing winds and thus may be offset from the systemic redshift by a few hundred km/s. The third column of the table gives the percentage of the total galaxy sample where each feature is detected at this step. Emission lines that are both faint and rare, such as the [O\textsc{iii}]4363 auroral line, require more careful attention to avoid false detections, and will be discussed in future work. 

This line list was shifted from $z=0$ to $z=2$ with step sizes of $\delta z = 0.001$. For each redshift, the residual flux in the spectrum was measured at the location of each redshifted line wavelength. If there was a peak with a signal-to-noise ratio (SNR) of 5 or more at the line wavelength, a detection was recorded, and after checking for each line, the SNRs of the significant peaks were combined. The redshift with the maximum cumulative SNR, and thus the most significantly detected lines, was selected as the redshift fit. We did not fit profiles to the emission lines at this stage, to avoid issues with lines with strong kinematic features or the semi-blended [O\textsc{ii}] doublet. If only one significant feature is detected, or if the SNR distribution has other peaks $\geq75\%$ the peak SNR, a second round of peak measurement occurs, checking for fainter features that are near to likely strong lines (e.g., the [S\textsc{ii}] doublet near the H$\alpha$ emission line).

This produces a first-round redshift fit, which is then visually examined and compared with the 2D spectra in order to flag false positives, such as the fitting of an unsubtracted skyline, or misidentifications, such as fitting a single [O\textsc{ii}] line as a single H$\alpha$ line. For this first round of fits, the spectra were rebinned by 10 pixels, in order to reduce the computation time over the larger redshift grid and to reduce the chances of fitting noisy spikes in the spectrum as emission lines. A second round of redshift fitting over a narrower range ($\Delta z =0.1$) is performed, incorporating any corrections from the visual inspection. For the second round, the spectra were left at their native resolution over a narrower redshift grid, allowing for more precise redshift measurements and the resolved detection of the [O\textsc{ii}] doublet. This has an expected redshift uncertainty of $\sigma_z = 0.001$. The signal threshold for line detection is lowered to 3 in order to include fainter but detectable emission lines (e.g., H$\gamma$).

After this second-round fit, we performed a final visual check of outliers and fits with a low number of emission line detections. Any remaining false detections or misidentifications were removed manually in this way. The redshift accuracy and redshift fit success rates are discussed in detail in \S \ref{sec:z_acc} and \S \ref{sec:sr}, where we find very close agreement with existing spectroscopic redshifts in the CANDELS fields.

\begin{deluxetable*}{cccccc}
\label{tab:z_class}
    \centering
    \tablecaption{Redshift Quality Flags} 
    \tablecolumns{6}
    \tablehead{
    \colhead{$z_Q$} &
    \colhead{Description} &
    \colhead{Definition} &
    \colhead{$N_{obj}$} &
    \colhead{$F_{obj}$(\%)} &
    \colhead{$\bar{t}_{exp}$}}
    \startdata
        3 & Secure & more than 2 strong features detected & 626 & 27 & 7.68 \\
        2 & Good & 2 strong features detected & 380 & 17 & 7.08 \\
        1 & Uncertain & only 1 strong feature detected & 426 & 19 & 6.16 \\
        0 & No fit & no significant spectral features detected & 858 & 37 & 6.16 \\
    \enddata
    %\tablenotetext{a}{The median exposure time of coadded spectra with this quality flag, in hours.}
    %\tablenotetext{b}{Possible strong features include the H$\alpha$,H$\beta$, and H$\gamma$ Balmer emission lines; the [O\textsc{iii}]4959, [O\textsc{iii} 5007], and semi-blended [O\textsc{ii}]3727,3729 emission lines; and the Ca H and K pair of absorption lines.}
    %\tablenotetext{c}{Includes objects rejected as stars, having excessive contamination in the spectra, etc.}
    
\end{deluxetable*}

\begin{figure*}
    \centering
    \includegraphics[width=\textwidth]{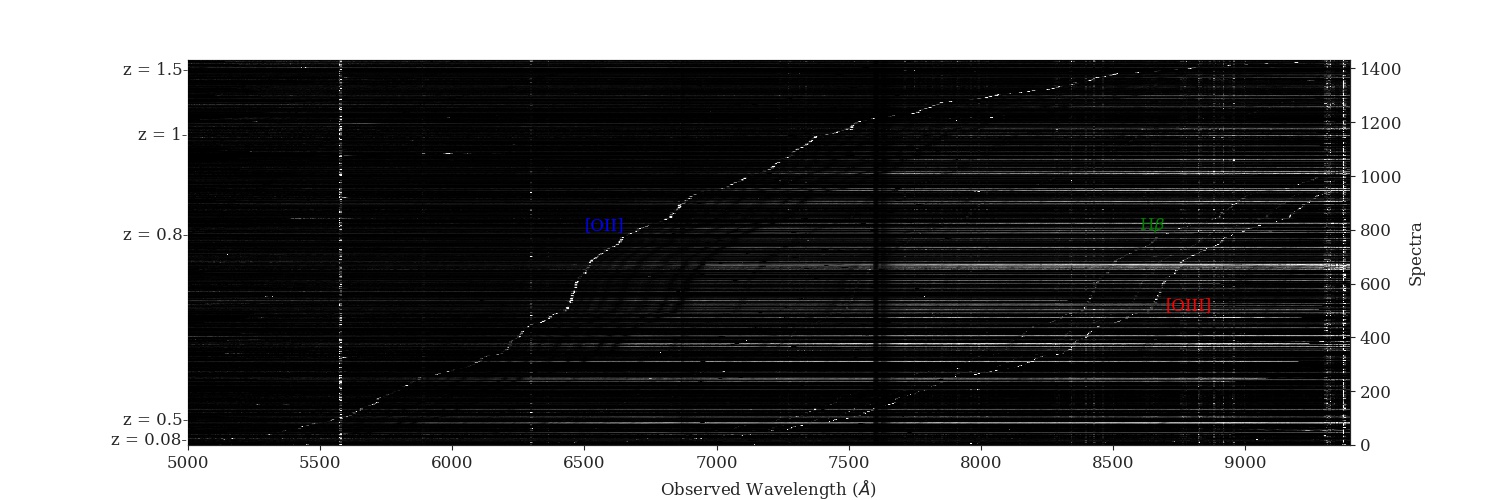}
    \caption{Overview of the HALO7D galaxies with redshift fits. The galaxies are sorted in ascending redshift from bottom to top. Note that the y-axis scale is not linear, as the sample is not uniformly distributed in redshift. The axis labels indicate a few example redshift locations to give a sense of this distribution. The brightness at each wavelength in each row is scaled to the flux of the DEIMOS spectrum.}
    \label{fig:elgs_arr}
\end{figure*}

With the final set of redshift fits, we classified the quality of the fits according to the number of strong features detected, which include the H$\alpha$, H$\beta$, and H$\gamma$ Balmer emission lines; the [O\textsc{iii}]4959, [O\textsc{iii} 5007] and [O\textsc{ii}]3727,3729 emission lines; and the Ca H and K pair of absorption lines. For this purpose we count a single strong emission line or the pair of Ca H and K absorption lines as a single feature. The [O\textsc{ii}]3727,3729 doublet is semi-blended at the resolution of the grating used, so its detection is counted as a single feature for the purpose of redshift classification. The classifications are summarized in Table \ref{tab:z_class}. The $z_Q = 0$ flag includes galaxies rejected as stars, having excessive contamination in the spectra, etc. as well as those with no data quality issues but nonetheless detect no strong features. The final three columns give the total number of galaxies with each flag, each flag's percentage of the total catalog, and the median exposure time of galaxies with a given flag.

We then present the redshift catalog, an excerpt of which is shown in Table \ref{tab:catalog}. The galaxy ID's are constructed in the form field\_h7ID, where `field' is an abbreviation of the CANDELS field name and h7ID is the ID name assigned within the HALO7D survey. This is the same as the CANDELS ID for those galaxies that match existing CANDELS catalogs. The full catalog also includes the galaxy ID's from COSMOS/UltraVista and EGS/IRAC where applicable. The $z_{H7}$ column gives the fit from HALO7D spectra, and the $z_{qual}$ column gives the quality flag described above. The Star column indicates whether the galaxy has been flagged as a star, either in matching catalogs or in our visual inspections of the spectra. The $z_{phot}$ and $z_{spec}$ columns give existing photometric and spectroscopic redshifts from matched catalogs \citep{dahlen13, muzzin13, barro19}, and are set to -99 if no such redshift is available. The Mass column gives the log of the stellar mass (as derived from SED fitting, see \S\ref{sec:dp}), and the F606W, F814W, and F160W columns give photometry in AB magnitudes. For a visual description of the redshift and emission line distributions, see Figure \ref{fig:elgs_arr}.

\begin{figure}
    \centering
    \includegraphics[width=0.5\textwidth]{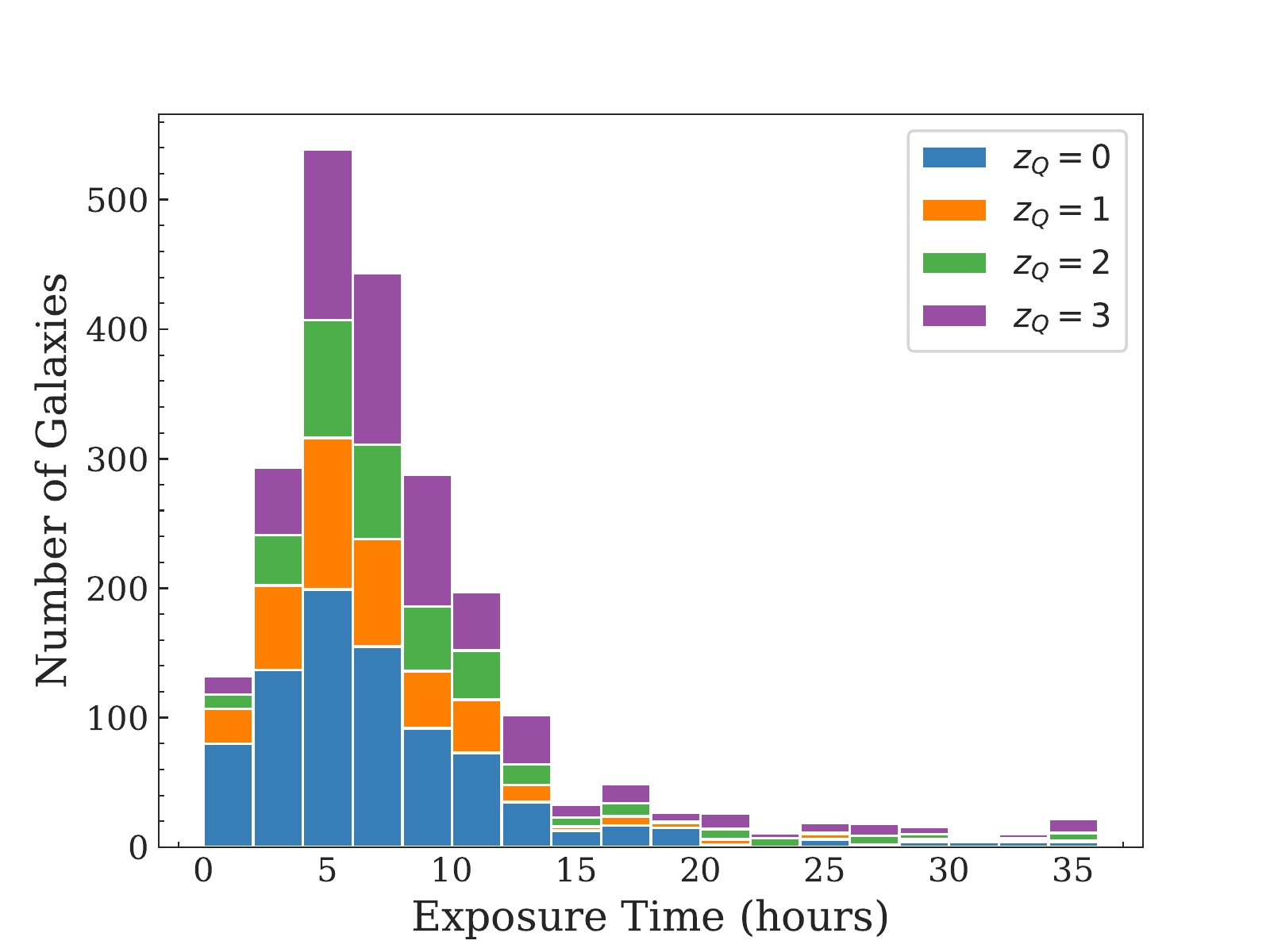}
    \caption{The distribution of total exposure time in the coadded spectra, measured in hours. The histogram is broken down by the assigned $z_Q$ redshift quality. This demonstrates the benefits of the depth of the observations, as the lowest-quality flag $z_Q=0$ is over-represented among galaxies with low exposure times, and the highest-quality flags are over-represented among the highest exposure times.}
    \label{fig:expt}
\end{figure}

Finally, the ExpTime columns gives the total exposure time in hours for the coadded HALO7D spectra. This total time excludes the time from exposures that were rejected from the coadd due to contamination or other issues. Figure \ref{fig:expt} gives the distribution of exposure times for the coadded spectra, colored by the redshift quality flag, and the median exposure time per quality flag is summarized in Table \ref{tab:z_class}. Higher $z_Q$ values are more likely to be assigned to spectra with high exposure times relative to lower $z_Q$ values, as would be expected if very deep exposures are needed to attain the signal necessary to detect spectral features in these galaxies. $z_Q$ values of 0 and 1 have a median exposure time of 6.16 hours, while $z_Q=2$ has a median time of 7.08 hours, and $z_Q=3$ has a median of 7.68 hours. The whole sample of spectra has a median of 6.59 hours. Note also that the $z_Q = 0$ flag includes objects rejected as stars, which explains the $z_Q=0$ spectra with large exposure times.

\begin{deluxetable*}{cccccccccccccc}
\centering
\label{tab:catalog}
\tablecaption{Redshift Catalog}
\tablecolumns{15}
\tablehead{
\colhead{ID} &
\colhead{$z_{H7}$} &
\colhead{$z_{qual}$} &
\colhead{Detected Lines} &
\colhead{RA} &
\colhead{Dec} &
\colhead{Star} &
\colhead{$z_{phot}$} &
\colhead{$z_{spec}$} & 
\colhead{LogMass} &
\colhead{F606W} &
\colhead{F814W} &
\colhead{F160W} &
\colhead{ExpTime} \\ 
\colhead{} &
\colhead{} &
\colhead{} &
\colhead{} &
\colhead{(Deg)} &
\colhead{(Deg)} &
\colhead{} &
\colhead{} &
\colhead{} &
\colhead{($M_{\odot}$)} &
\colhead{(Mag)} &
\colhead{(Mag)} &
\colhead{(Mag)} &
\colhead{(hours)} \\
}
\rotate
\startdata
%1 & 1 & 1 & 1 & 1 & 1 & 1 & 1 & 1 & 1 & 1 & 1 & 1 & 1 & 1 \\
cos\_10046u & 0.942 & 3 & O\textsc{ii},CaK,CaH,H$\gamma$ & 150.0015 & 2.4845 & 0 & 0.948 & 0.9414 & 10.58 & -99.0 & -99.0 & -99.0 & 7.08 \\
cos\_10050 & 0.517 & 2 & O\textsc{ii},CaH,CaK & 150.18735 & 2.2991278 & 0 & 0.51 & -99 & 9.489 & 23.17 & 22.37 & 21.37 & 8.35 \\
cos\_10064 & -99 & 0 & none & 150.13917 & 2.2998361 & 0 & 0.08 & -99 & 7.367 & 23.08 & 22.83 & 22.66 & 7.02 \\
\enddata
\end{deluxetable*}

\begin{deluxetable*}{cccccccc}
\centering
\label{tab:flux}
\tablecaption{Emission Line Flux Catalog}
\tablecolumns{8}
\tablehead{
\colhead{ID} &
\colhead{$z_{H7}$} &
\colhead{ELC\tablenotemark{a}} &
\colhead{Detected Lines} &
\colhead{[O\textsc{ii}] Flux\tablenotemark{b}} & %\tablenotemark{c} &
\colhead{[O\textsc{ii}] Error} & %\tablenotemark{c}  &
\colhead{[O\textsc{ii}] SNR} &
\colhead{[O\textsc{ii}] EW} \\
%\colhead{[O\textsc{ii}]5007 Flux}\tablenotemark{b}  &
%\colhead{[O\textsc{ii}]5007 Err}\tablenotemark{b}  &
%\colhead{[O\textsc{ii}]5007 SNR} &
%\colhead{[O\textsc{ii}]5007 EW} \\
\colhead{} &
\colhead{} &
\colhead{} &
\colhead{} &
\colhead{(erg cm$^{-2}$ s$^{-1}\times10^{19}$)} &
\colhead{(erg cm$^{-2}$ s$^{-1}\times10^{19}$)} &
\colhead{} &
\colhead{(\AA)} %&
%\colhead{(erg cm$^{-2}$ s$^{-1}$)} &
%\colhead{(erg cm$^{-2}$ s$^{-1}$)} &
%\colhead{} &
%\colhead{(\AA)}
}
%\rotate
\startdata
cos\_10046u & 0.942 & 2 & [O\textsc{ii}],H$\gamma$ & 22.56 & 1.87 & 12.08 & 23.92 \\
cos\_10050 & 0.517 & 3 & [O\textsc{ii}],H$\beta$,[O\textsc{iii}]5007 & 537.38 & 25.82 & 20.81 & 53.02 \\
cos\_10064 & -1.0 & 0 & none & -99.0 & -99.0 & -99.0 & -99.0  \\
cos\_10105u & -1.0 & 0 & none & -99.0 & -99.0 & -99.0 & -99.0  \\
cos\_10122 & 0.94 & 1 & [O\textsc{ii}] & 301.75 & 21.64 & 13.95 & 12.74  \\
cos\_10191 & -1.0 & 0 & none & -99.0 & -99.0 & -99.0 & -99.0  \\
%cos\_1022 & 0.683 & 4 & [O\textsc{ii}],H$\beta$,[O\textsc{iii}]4959,[O\textsc{iii}]5007 & 328.2 & 21.91 & 14.98 & 7.27  \\
%cos\_1029 & 0.728 & 6 & [O\textsc{ii}],H$\delta$,H$\gamma$,H$\beta$,[O\textsc{iii}]4959,[O\textsc{iii}]5007 & 1374.56 & 56.55 & 24.31 & 91.85  \\
\enddata
\tablenotetext{a}{The Emission Line Count, the number of significantly detected emission lines for which fluxes are measured. May differ from the counts in the redshift catalog, as fainter lines are included.}
\tablenotetext{b}{This column contains the summed flux of the [O\textsc{ii}]3727,3729 doublet}
%\tablenotetext{c}{Reported fluxes and errors are $\times 10^{19}$}
\end{deluxetable*}

\begin{figure*}
    \centering
    \includegraphics[width=\textwidth]{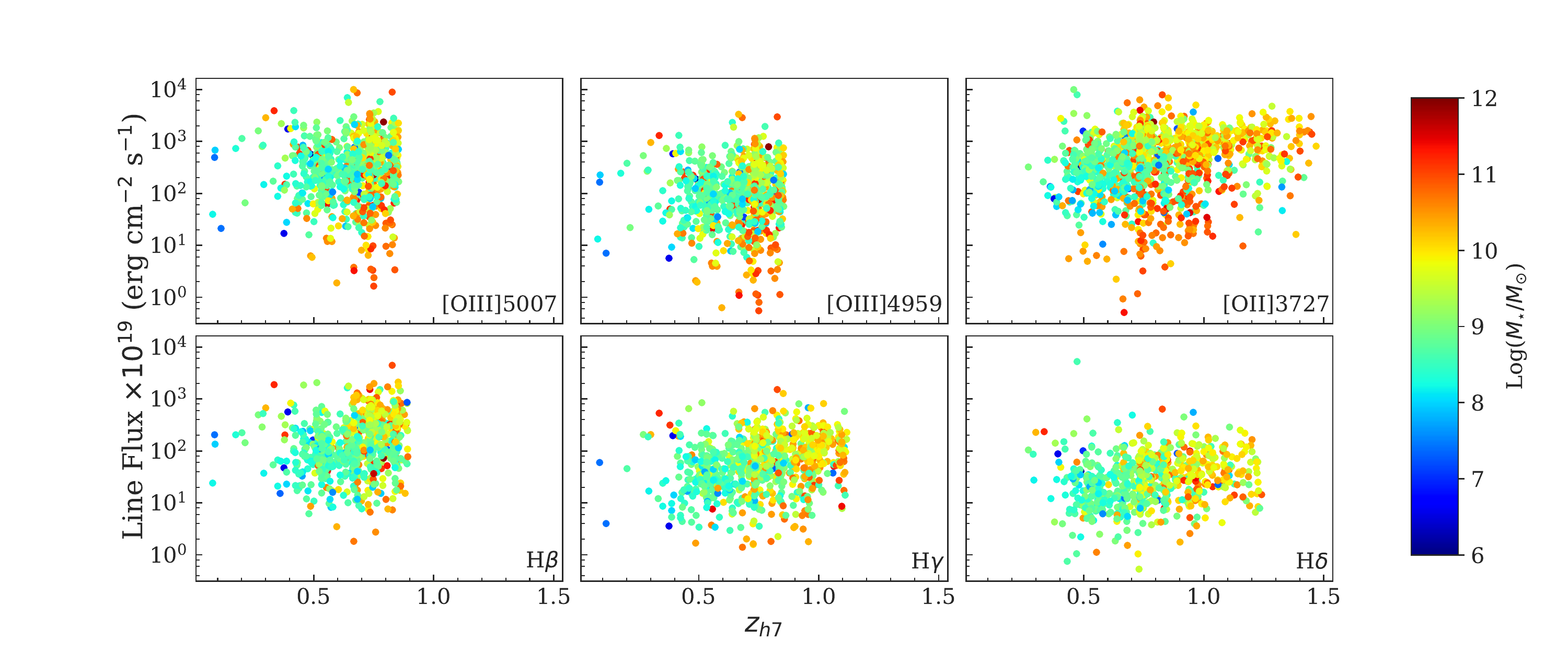}
    \caption{The emission line flux distributions as a function of redshift, and colored by the stellar mass. The top row gives ionized oxygen emission lines, and the bottom row hydrogen Balmer series emission. Fluxes are not corrected for dust extinction or stellar absorption.}
    \label{fig:el_fluxes}
\end{figure*}

\subsection{Line Flux and Equivalent Width} \label{subsec:flew}

With the final redshifts determined, we next measured the emission line fluxes. After removing the estimated continuum flux (as described in \S\ref{subsec:z}), the total line flux and flux error is measured by integrating the continuum-subtracted line flux region. This integration is bounded between the points to the left and right of the line center (as defined by the spectroscopic redshift) where the line flux rejoins the continuum, which in the continuum-subtracted spectrum are the points where the residual flux reaches 0. This method avoids potential errors from assuming an incorrect line shape, and is consistent with fluxes derived from lines well-fit by a single Gaussian. The ratio of this integrated line flux and the local continuum estimate gives the line equivalent width (EW).

Any line for which the ratio of the total line flux to the line error is $\geq3$ is recorded as a detection. See Table \ref{tab:flux} for an excerpt of the line flux catalog. For the sake of space, only the combined [O\textsc{ii}] line columns are shown.

The distribution of the emission lines fluxes are shown in Figure \ref{fig:el_fluxes}, and the median properties of the emission lines are given in Table \ref{tab:el_stats}. The most commonly detected line is the [O\textsc{ii}] doublet, as it falls within the DEIMOS spectroscopic coverage for a large range of redshift, as it is functionally detectable for $0.3 < z < 1.5$ and it may be strong in both the low- and high-mass galaxy samples. [O\textsc{iii}] and H$\beta$ are preferentially detected at somewhat lower redshift and lower stellar mass, due to a combination of a more limited redshift range and a required level of ionization that is more common in star-forming galaxies, while much of the massive sample is quiescent. Though there are a handful of H$\alpha$ line detections included in the catalog, the line is too red at these redshifts to be commonly detected.

The non-uniform selection of target galaxies and the variable exposure times mean the survey sensitivity may not be uniform. However, Figure \ref{fig:el_fluxes} does demonstrate detections down to line fluxes of $10^{-19}$ erg cm$^{-2}$ s$^{-1}$. Such detections are almost entirely at $z < 1$. At higher redshifts, the selected targets are primarily massive galaxies with strong stellar continua, for which lines of this faintness would have very low EWs, making their detection extremely difficult. The distributions of [O\textsc{ii}], H$\delta$, and H$\gamma$ emitters do suggest that we are able to detect lines down to fluxes of $10^{-18}$ erg cm$^{-2}$ s$^{-1}$ across all redshifts in the survey.

\begin{deluxetable*}{cccccccc}
\centering
\tablecaption{Median Emission Line Properties}
\tablecolumns{6}
\tablehead{
\colhead{Line} &
\colhead{$\lambda_{\text{rest}}$} &
\colhead{$z$} &
\colhead{$N$} &
\colhead{$N_{dwarf}$} &
\colhead{Flux\tablenotemark{{a}}} &
\colhead{Stellar Mass} &
\colhead{EW$_{\text{obs}}$} \\
\colhead{} &
\colhead{\AA} &
\colhead{} &
\colhead{} &
\colhead{} &
\colhead{} &
\colhead{Log($\frac{M_{\star}}{M_{\odot}}$)} &
\colhead{\AA}}
\startdata
[O\textsc{ii}] & 3727,3729 & 0.767 & 1305 & 584 & 44 & 9.67 & 62 \\
H$\delta$ & 4102 & 0.751 & 679 & 403 & 3 & 9.23 & 3 \\
H$\gamma$ & 4341 & 0.745 & 875 & 506 & 6 & 9.20 & 6 \\
H$\beta$ & 4861 & 0.705 & 732 & 484 & 14 & 8.98 & 19 \\
$\left[ \text{O}\textsc{iii} \right]$\tablenotemark{b} & 5007 & 0.696 & 810 & 493 & 31 & 9.04 & 39 
\enddata
\tablenotetext{a}{Flux given in units of $10^{-18}$ erg cm$^{-2}$ s$^{-1}$.}
\tablenotetext{b}{This line was detected simultaneously with [O\textsc{iii}]4959, which has the same median $z$ and mass.}
\label{tab:el_stats}
\end{deluxetable*}

\section{Redshift Properties} \label{sec:red}

\subsection{Redshift Accuracy} \label{sec:z_acc}

\begin{deluxetable}{cccc}
\centering
\tablecaption{Comparison with Existing Spectroscopic Redshifts}
\tablecolumns{3}
\tablehead{
\colhead{Description} &
\colhead{} &
\colhead{$N_{obj}$} &
\colhead{$F_{obj}$(\%)}}
\startdata
$z_{h7}$ Fits & & 1440 & 63 \\ \hline
& $z_{spec}$ Agreement\tablenotemark{a} & 601 & 26 \\
& $z_{spec}$ Outlier\tablenotemark{a} & 21 & 1 \\
& High $z_{spec}$ & 8 & $<1$ \\
& New $z$ fit & 810 & 35 \\ \hline
No $z$ fit & & 850 & 37 \\ \hline
& Stars & 102 & 4 \\
& $z_{spec}$, no $z_{h7}$ & 50 & 2 \\
& No $z$ & 698 & 30 \\ \hline
Total Catalog & & 2290 & 100 \\
\enddata
\tablenotetext{a}{$z_{spec}$ agreement is defined as having $|(z_{h7}-z_{spec})|/(1+z_{spec}) < 0.01$; otherwise the fit is marked as an outlier.}
\label{tab:z_comp}
\end{deluxetable}

To gauge the accuracy of our redshift fits, we compare the results with a catalog of existing spectroscopic redshifts compiled from previous CANDELS data products (see \S\ref{sec:dp}. The results of this comparison are summarized in Table \ref{tab:z_comp}. For those galaxies with existing spectroscopic redshifts (spec-z), our results closely match the spec-z ($|z_{spec} - z_{HALO7D}| /(1+z_{spec}) < 0.01$) for 601 galaxies, with 21 outliers. There are 58 galaxies with existing spectroscopic redshifts where we are unable to obtain a fit with HALO7D spectra. For 8 of the latter galaxies, visual inspection suggests the presence of a high-z emission line (e.g., Mg\textsc{ii} 2800\AA) that was not included in the fitting procedure but can be identified visually and thus are included in the final catalog. The other 50 are primarily higher-mass galaxies with older galaxy populations, likely fit from absorption lines or stellar continua, which are not part of our analysis but will be explored in other HALO7D studies. We obtain good fits for an addition 810 galaxies without existing spec-zs, for a total of 1440 galaxies with at least 1 line detection. This includes 455 galaxies with a stellar mass of $M_{\star} < 10^{9.5} M_{\odot}$ for which we provide a new redshift fit.

\begin{figure}
    \centering
    \includegraphics[width=0.5\textwidth]{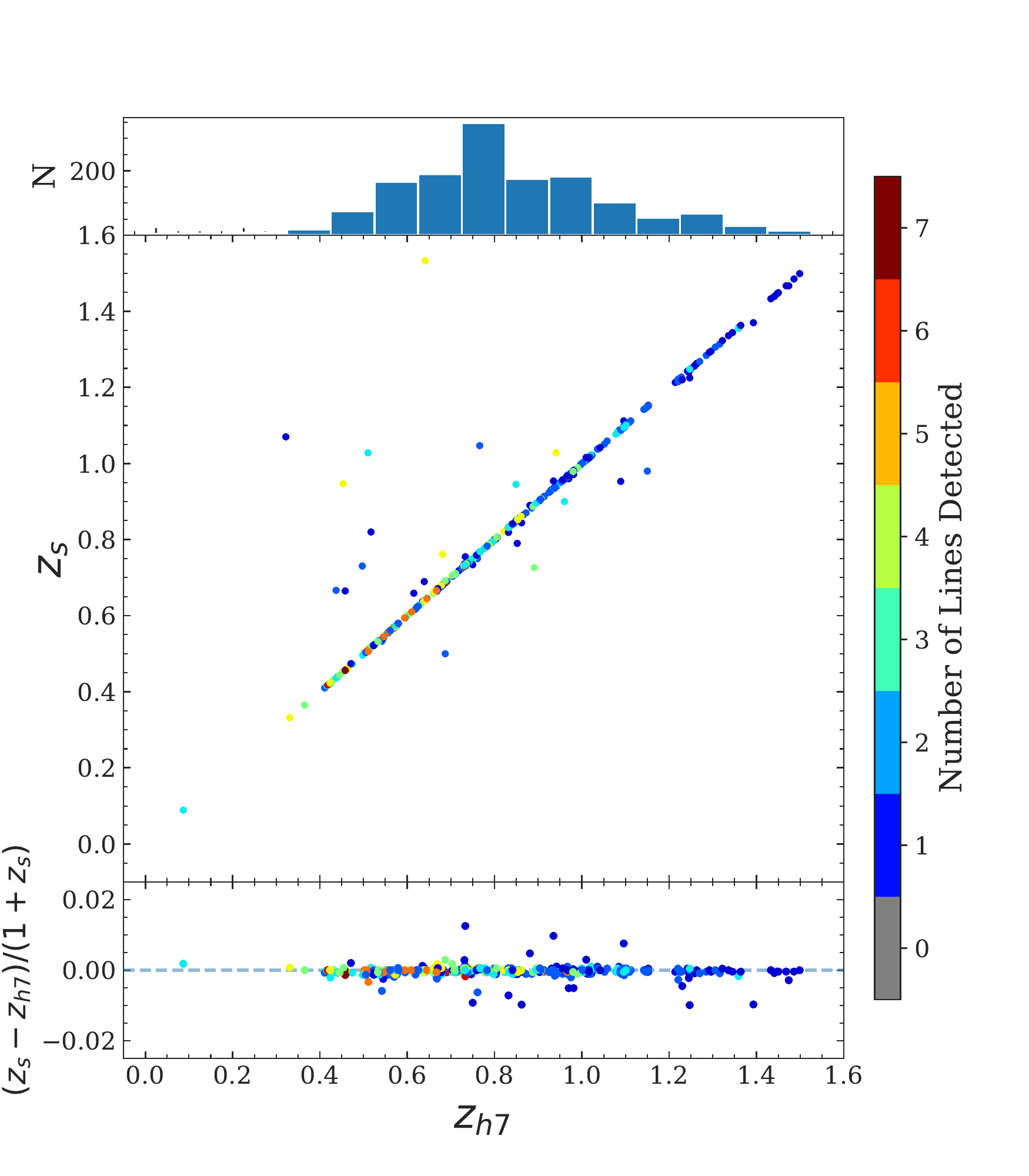}
    \caption{\textit{Top panel:} Histogram of $z_{h7}$, redshifts from this work where there is a good fit with $z > 0$. \textit{Middle panel:} other catalog spectroscopic redshifts $z_s$ compared to $z_{h7}$, colored according to the number of features detected. Galaxies without spectroscopic redshifts from other catalogs are set to $z_s = 0$, and galaxies without a good fit from HALO7D spectra are set to $z_{h7} = 0$. \textit{Bottom:} Redshift difference between $z_s$ and $z_{h7}$, normalized by $1+z_s$.}
    \label{fig:z_comp}
\end{figure}

Figure \ref{fig:z_comp} shows a direct comparison of the HALO7D redshift fits with existing spec-zs, taken from the compilations in \citet{dahlen13, muzzin13, barro19}. 

\subsection{Success Rate} \label{sec:sr}

\begin{figure*}
\centering
\begin{tabular}{cc}
    \includegraphics[width=0.45\textwidth]{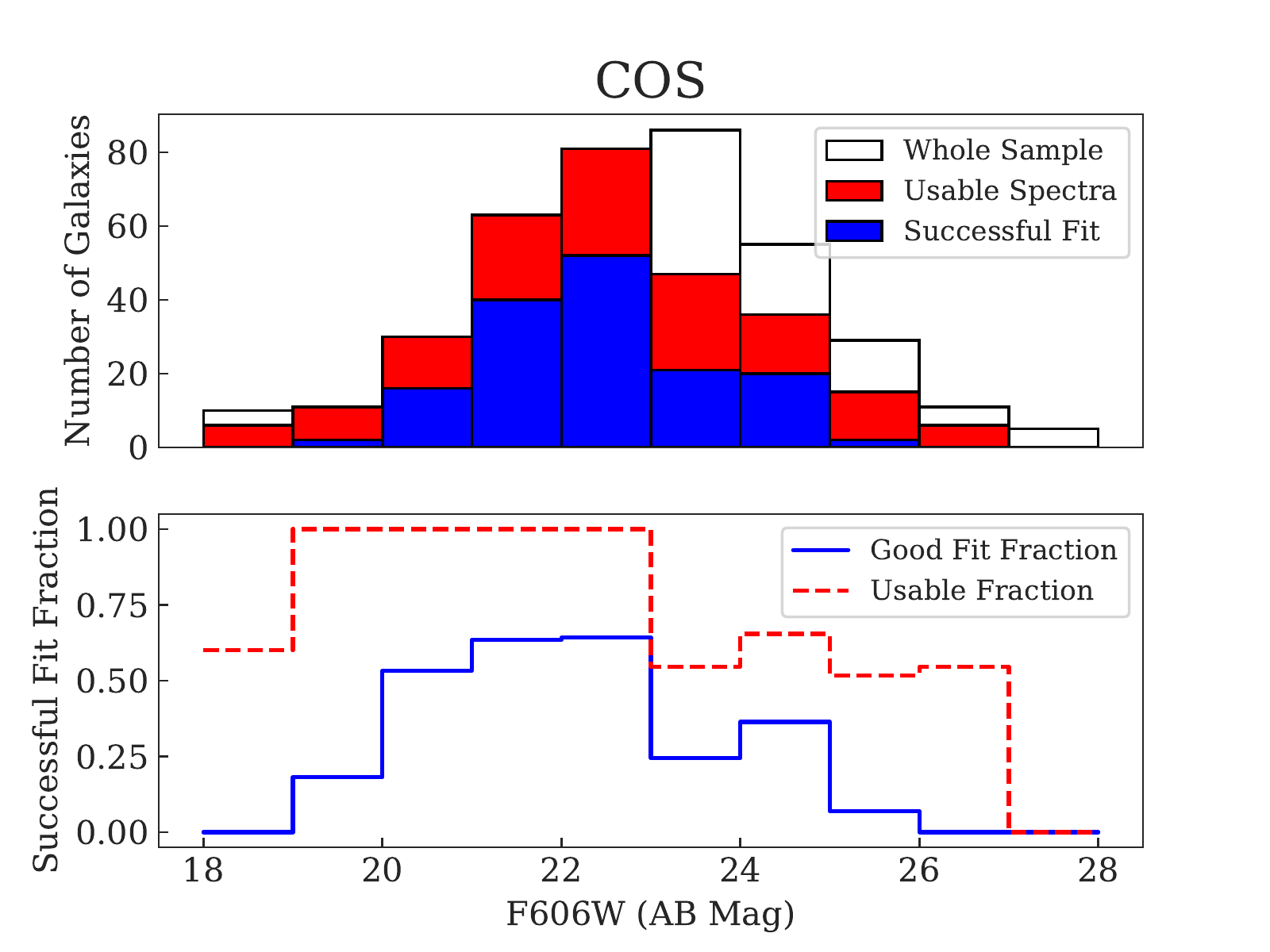} & \includegraphics[width=0.45\textwidth]{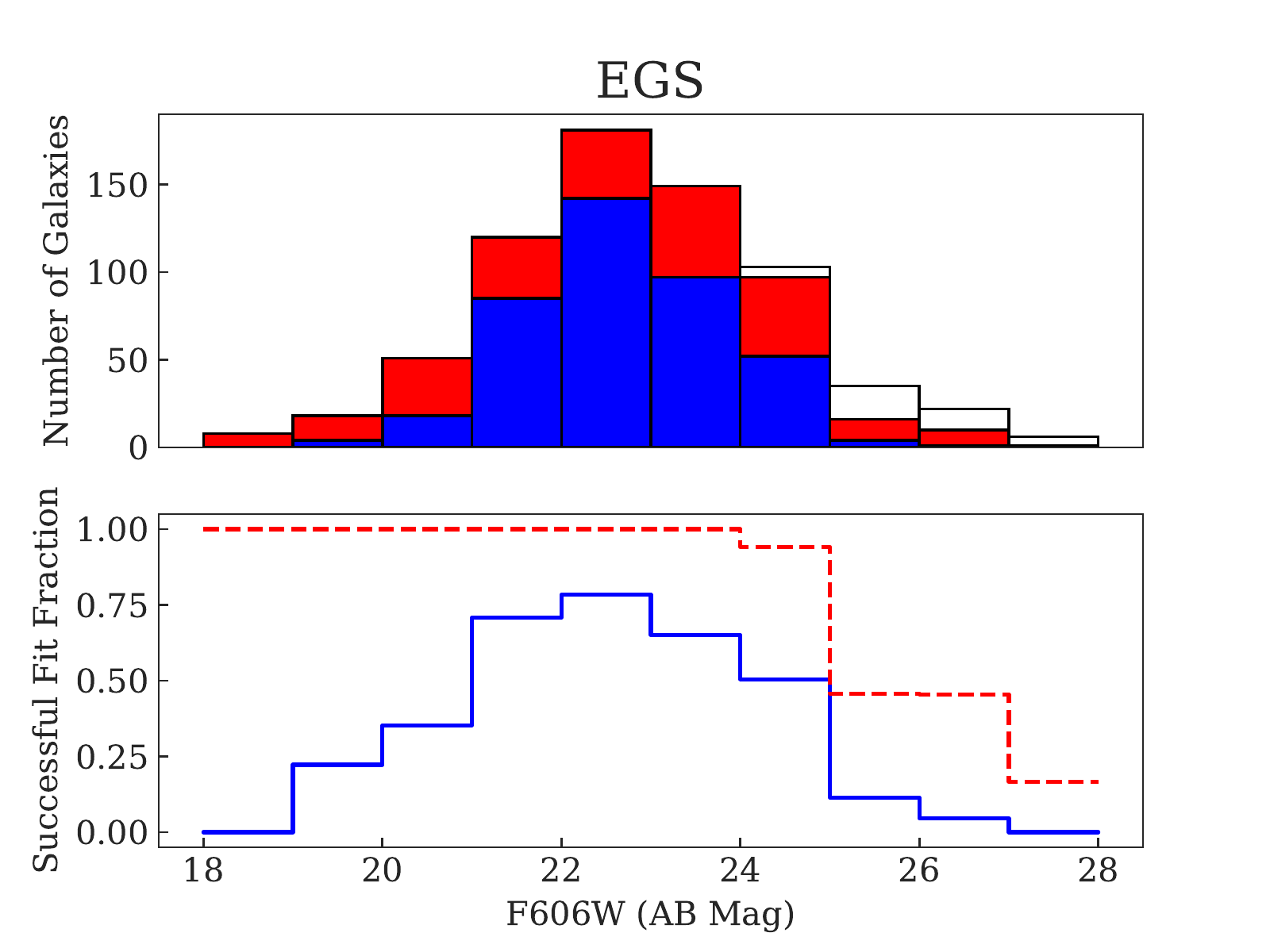} \\
    \includegraphics[width=0.45\textwidth]{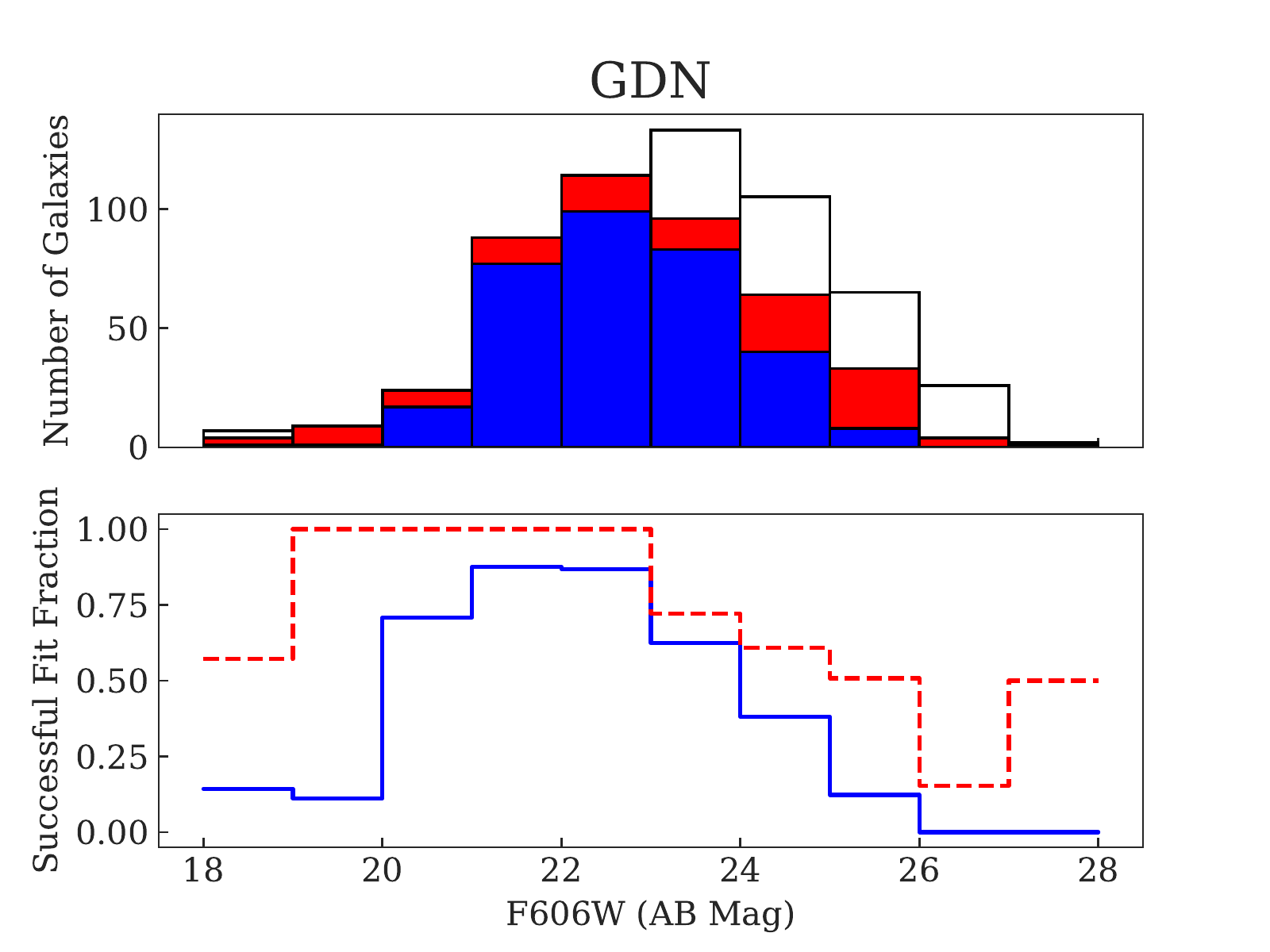} & \includegraphics[width=0.45\textwidth]{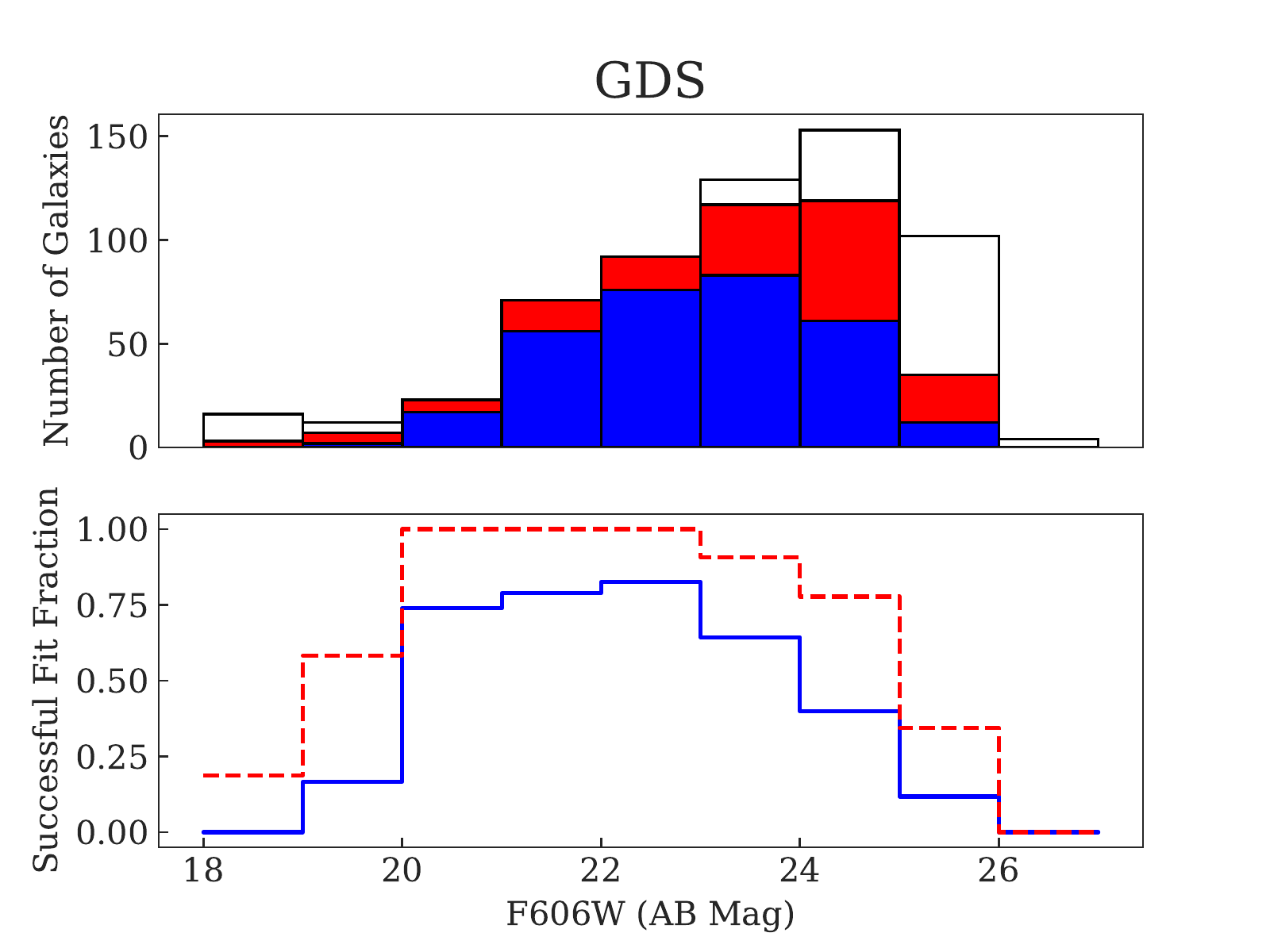}
\end{tabular}
\caption{Each pair of panels gives the F606W AB magnitude histograms and success fractions for one of the four HALO7D fields. The top panels give magnitude histograms for the whole target sample (white), the sample with usable extracted spectra (red), and the sample with good redshift fits (blue). The bottom panels give the success fractions as a function of magnitude. The red dashed line is the fraction of usable spectra out of the target sample in a given magnitude bin. The solid blue line gives the fraction of spectra with good redshift fits out of the total target sample.}
\label{fig:phot_success}
\end{figure*}

\begin{figure*}
\centering
\begin{tabular}{cc}
    \includegraphics[width=0.45\textwidth]{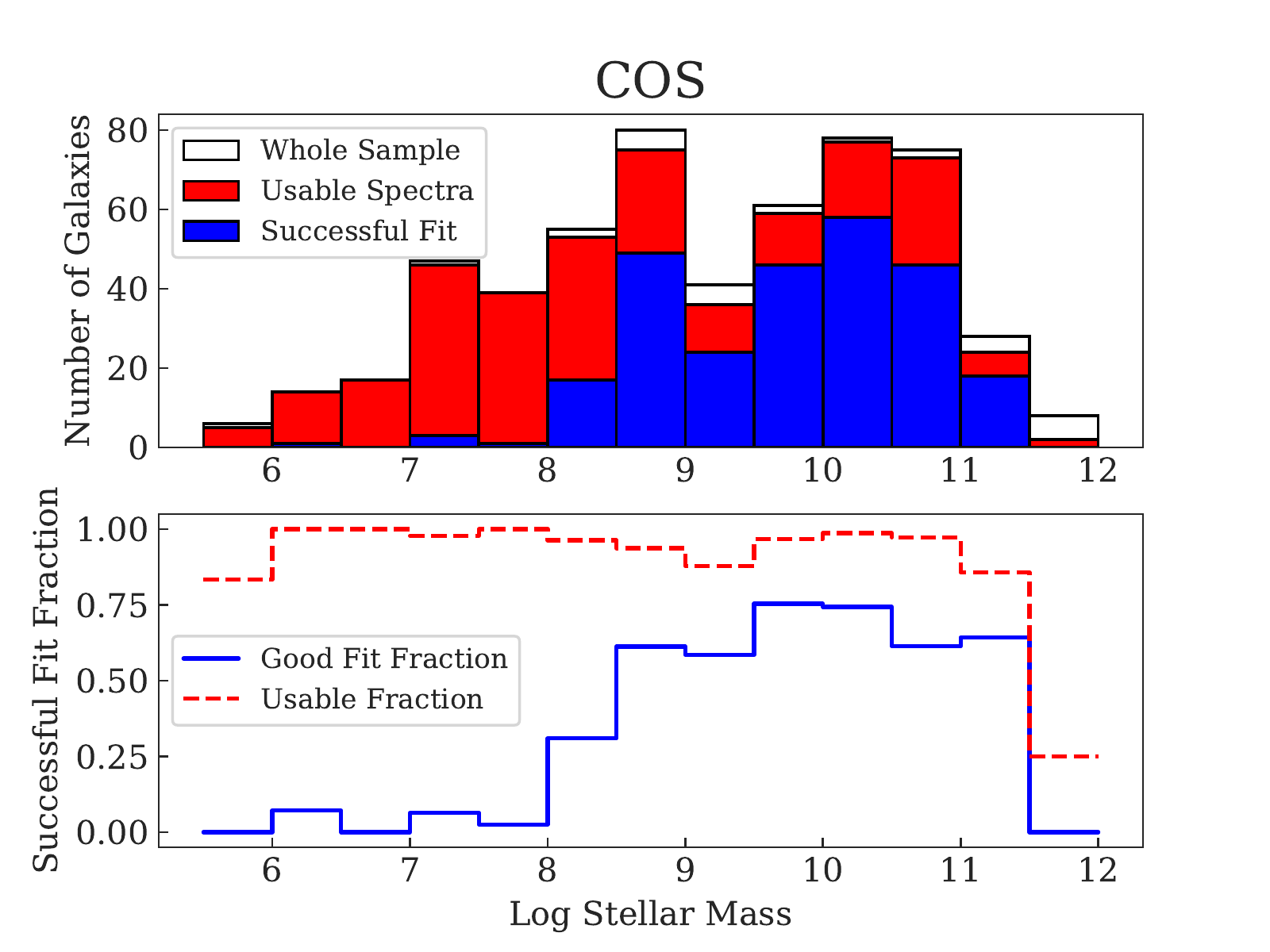} & \includegraphics[width=0.45\textwidth]{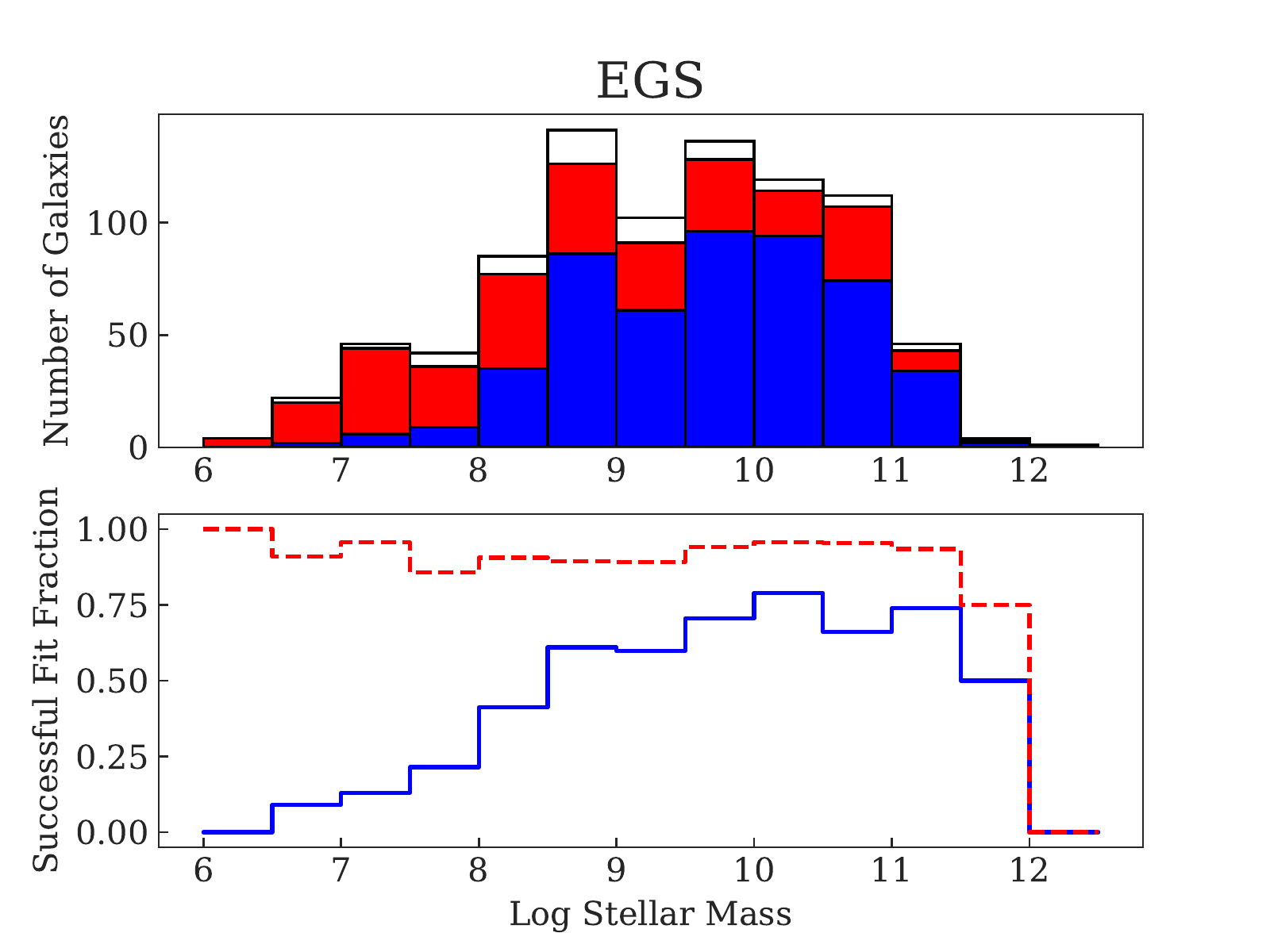} \\
    \includegraphics[width=0.45\textwidth]{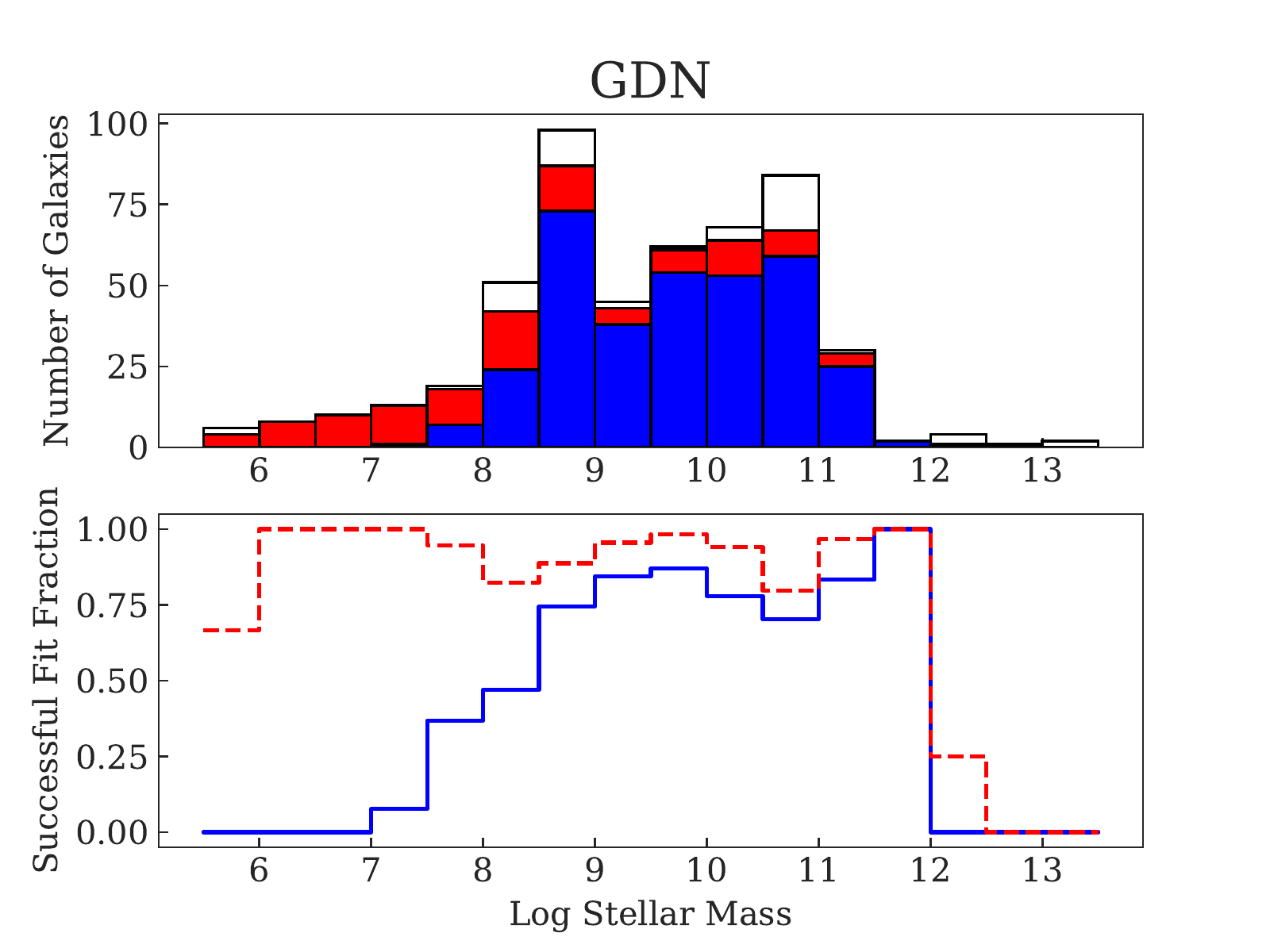} & \includegraphics[width=0.45\textwidth]{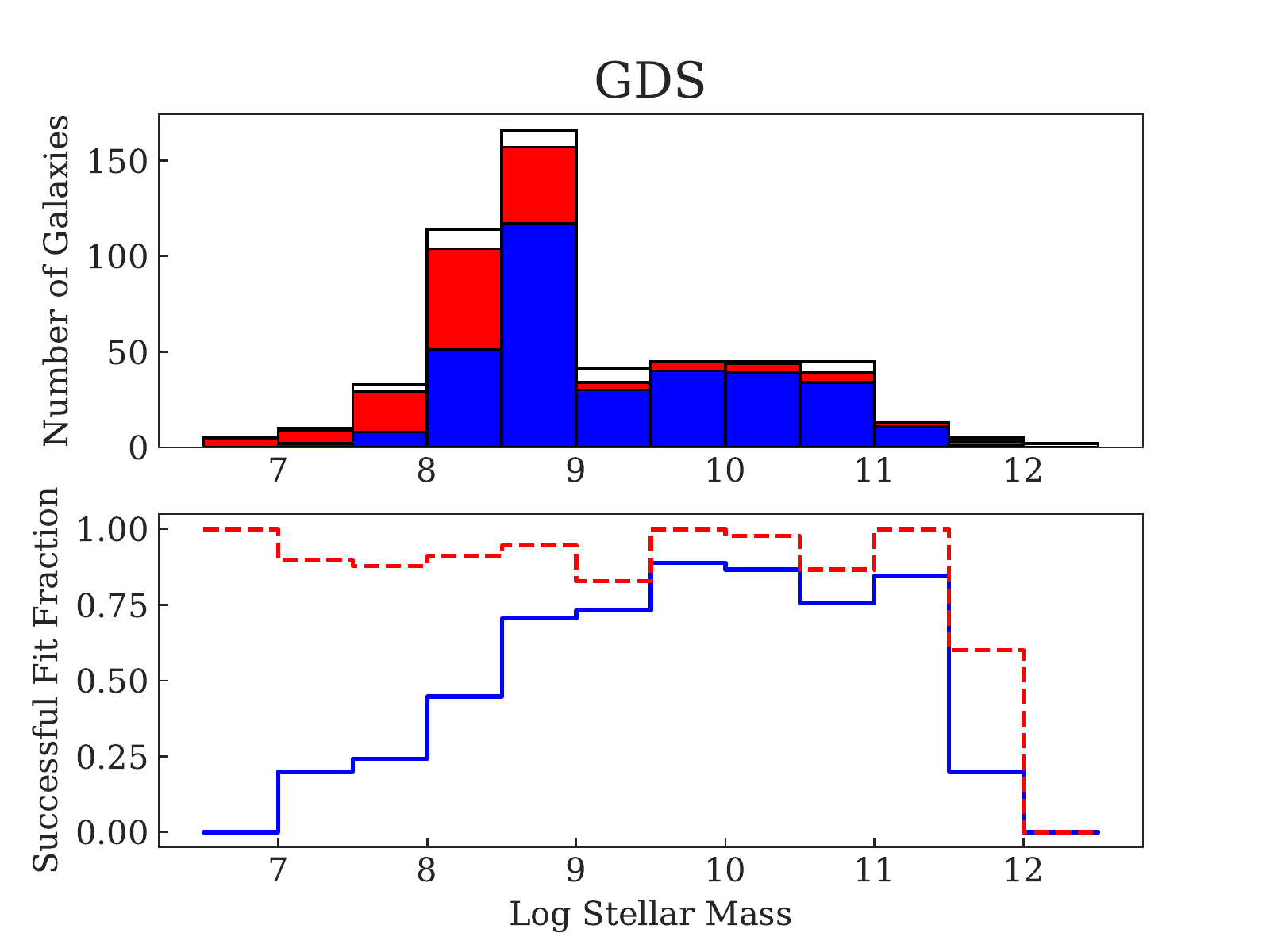}
\end{tabular}
\caption{Each pair of panels gives the stellar mass histograms and success fractions for one of the four HALO7D fields. The top panels give mass histograms for the whole target sample (white), the sample with usable extracted spectra (red), and the sample with good redshift fits (blue). The bottom panels give the success fractions as a function of stellar mass. The red dashed line is the fraction of usable spectra out of the target sample in a given mass bin. The solid blue line gives the fraction of spectra with good redshift fits out of the total target sample.}
\label{fig:mass_success}
\end{figure*}

In this section, we investigate the fraction of HALO7D galaxies for which we obtained successful redshift fits, measured as a function of field, magnitude, and stellar mass. HALO7D target galaxies are drawn from several surveys (see \S\ref{sec:dp}), which target different fields and have different selection functions by redshift, stellar mass, and magnitude. We thus analyze the successful fit fraction separately by field to account for this. 

Figure \ref{fig:phot_success} shows successful fit rates as a function of the F606W magnitude (approximately the blue chip continuum magnitude for the spectrum), with a separate panel for each field. In each field, the top panel gives overlapping histograms of galaxies in 1-mag bins. The white histogram is the distribution of the total target sample, while red gives the distribution of galaxies with usable spectra after visual inspection. The blue histogram gives the distribution of successful redshift fits ($z_Q > 0$). The lower panels give the rate of usable spectra per magnitude (red dashed line) and the rate of successful redshift fit (solid blue line).

We typically find success rates of $\geq 50\%$ down to F606W$< 24$ mag, with the successful fits then declining down to negligible at 26 mag and fainter. We measure lower success rate in the COSMOS field, though this could be explained by the relatively large number of extremely low-mass ($\log(M_{\star}/M_{\odot}) \leq 8$) targets in that field. Overall, our success rates are comparable to the success rate shown in DEIMOS spectra from the Team Keck survey of GOODS-North (GDN) \citep{wirth04}, though that survey has a limiting \textit{R}-band magnitude of 24.4, while this work achieves some successful fits at F606W $> 25$ mag. Low success rates at high magnitudes are due to the flagging and removal of stars in the sample, as well as a low overall number of such objects.

Figure \ref{fig:mass_success} shows the successful fit fraction as a function of the log of the stellar mass, with the same organization and coloring as Figure \ref{fig:phot_success}. For masses $ \log(M_{\star}/M_{\odot}) > 9.5$, which we will classify as high-mass galaxies, we achieve success rates of $\sim75\%$ across all four fields. For low-mass galaxies, where $\log(M_{\star}/M_{\odot}) < 9.5$, we have an overall successful fit fraction of 49\%. However, galaxies with $8.5 < \log(M_{\star}/M_{\odot}) < 9.5$ are fit successfully at rates comparable to the higher mass galaxies, with successful fits dropping substantially at $\log(M_{\star}/M_{\odot}) < 8.5$. This suggests we are able to measure galaxies with $8.5 < \log(M_{\star}/M_{\odot}) < 9.5$ at a level of sensitivity comparable to the high-mass sample, with differences in completeness then arising only from differences in target selection as a function of mass. With this level of completeness, the HALO7D catalog provides a robust low-mass galaxy sample for studying trends in galaxy properties as a function of mass.

\subsection{Comparison with Photometric Redshifts}

\begin{figure}
    \centering
    \includegraphics[width=0.5\textwidth]{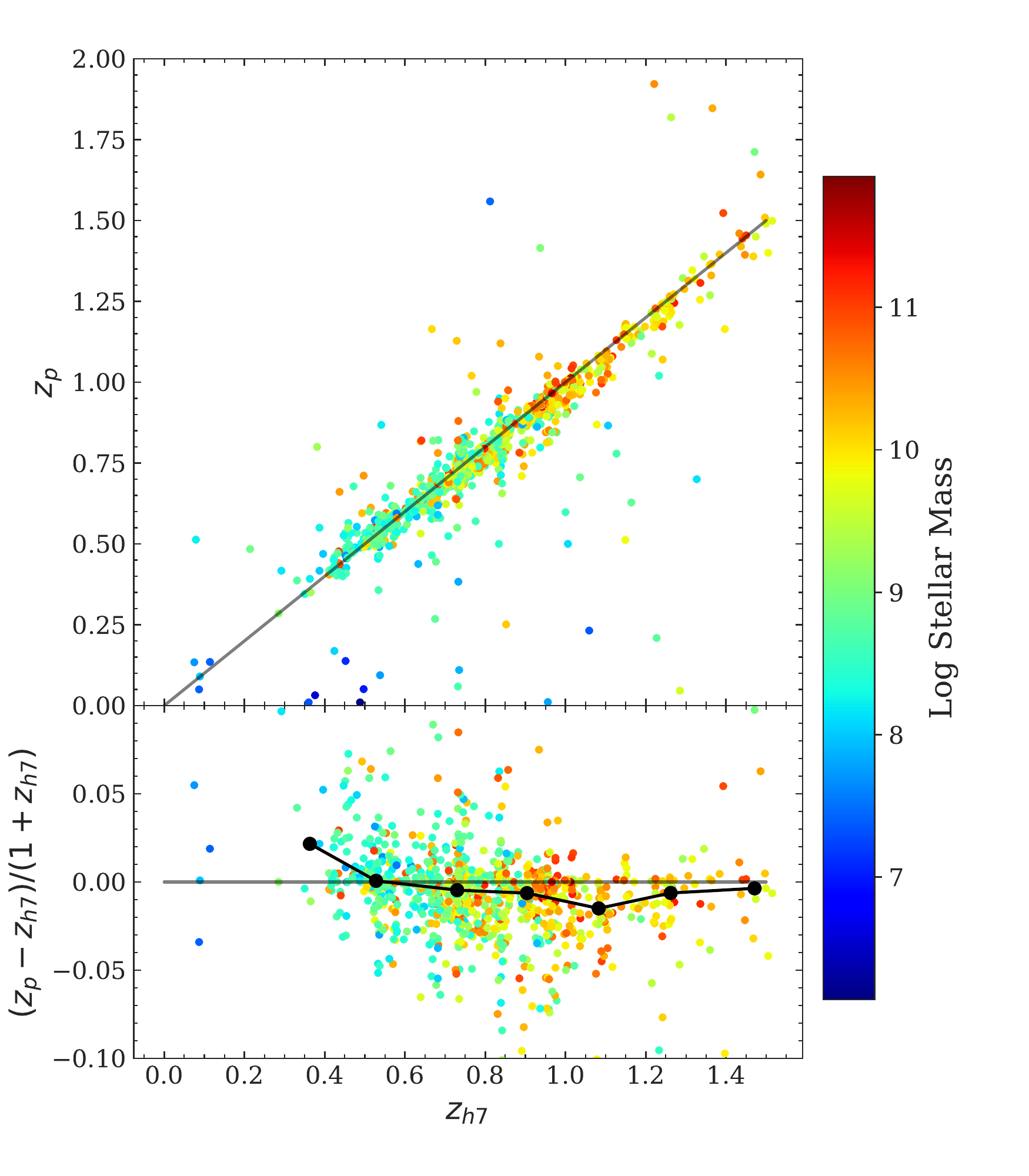}
    \caption{\textit{Top:} Photometric redshifts $z_p$ (primarily from CANDELS \citep{dahlen13}, EGS/IRAC \citep{barro11a,barro11b}, and COSMOS/UltraVista \citep{muzzin13}) compared to the redshift from HALO7D spectra $z_{h7}$, colored by the stellar mass. \textit{Bottom:} The difference in redshift between $z_{h7}$ and $z_p$, colored by the stellar mass. The solid black points show the median difference as a function of $z_{h7}$.}
    \label{fig:zphot_mass}
\end{figure}

We have also compared our redshifts with photometric redshifts from the CANDELS \citep{dahlen13,guo13,santini15,kocevski17,barro19}, COSMOS/UltraVista \citep{muzzin13}, and EGS/IRAC \citep{barro11a,barro11b} catalogs. The CANDELS photometric redshifts make up 80\% of the combined photometric redshift sample. The catalog assembled in \citet{dahlen13} is composed of redshift fits derived from combinations of results from 11 different photometric redshift fitting codes, eight of which accounted for line emission in their galaxy templates. They found that the combination of results from a range of codes reduced the photometric redshift scatter relative to high-quality spectroscopic redshifts. Figure \ref{fig:zphot_mass} supports this, showing generally good agreement between HALO7D spectroscopic redshifts and photometric redshifts. Photometric redshifts included from other catalogs were derived with the EAZY \citep{brammer08}, LePhare \citep{ilbert2006}, and WikZ \citep{wiklind2008} codes, of which EAZY alone includes emission lines in templates.

For $\Delta z = z_p - z_{H7}$, we find a median $|\Delta z| /(1+z_{H7})$ of 0.011 for the entire sample, with 25th ($Q_1$) and 75th ($Q_3$) percentile values of 0.004 and 0.026. However, the lower panel in Figure \ref{fig:zphot_mass} does indicate a tendency for the photometric redshifts to underestimate the spectroscopic redshift, particularly at around $1 < z < 1.2$. HALO7D galaxies at this redshift are most likely identified by [O\textsc{ii}] emission. If the photo-z galaxy templates don't anticipate the strength of the [O\textsc{ii}] line, it may interpret the relative drop in flux in the broadband to the blue of [O\textsc{ii}] as corresponding with the characteristic drop in flux associated with the 4000\AA\ break, which would underestimate the redshift somewhat. 

\begin{figure*}
    \centering
    \includegraphics[width=\textwidth]{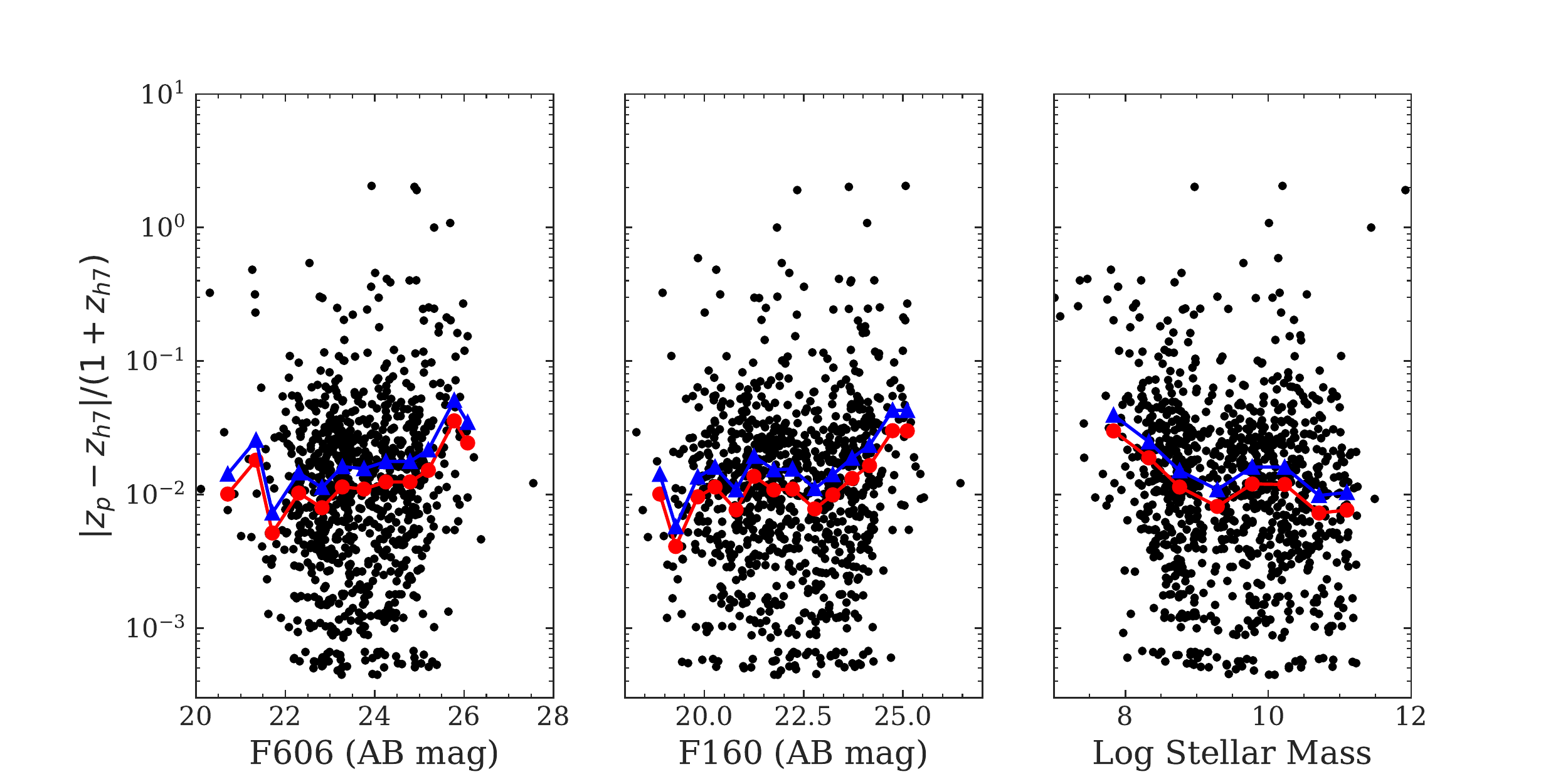}
    \caption{The absolute difference between the photometric redshift $z_p$ and the HALO7D redshift $z_{h7}$, plotted against \textit{Left:} the F606W AB magnitude, \textit{Middle:} the F160W AB magnitude, and \textit{Right:} the stellar mass. Red points give the median absolute difference in bins of magnitude or mass, and blue points give the $\sigma_{NMAD}$.}
    \label{fig:zphot}
\end{figure*}

In Figure \ref{fig:zphot}, we investigate $\Delta z$ as functions of F606W and F160W magnitude and stellar mass. We find small increases in $\Delta z$ and $\sigma_{NMAD}$\footnote{Defined using the form described in \citep{brammer08}, where $\sigma_{NMAD} = 1.48 \times \text{median} \left( \left| \frac{\Delta z - \text{median}(\Delta z)}{(1+z_{H7})} \right| \right)$} with increasing magnitude and decreasing stellar mass. The median redshift error only increases slightly with lower mass, rising from $|\Delta z| /(1+z_{H7}) = 0.010$ to 0.013 from the high-mass to low-mass bins. The 25h and 75th percentile values change from $Q_1 = 0.0036$ to $Q_1 = 0.0045$ and $Q_3=0.023$ to $Q_3=0.032$.

This can be seen in Figure \ref{fig:zphot_mass} as well, where the points are colored by stellar mass, and there is somewhat higher scatter at lower redshifts, where the lower-mass detections dominate. However, the lower panel from Figure \ref{fig:zphot_mass} suggests there is not much of a systematic error: while the absolute error increases for lower masses, the median $\Delta z$ values remain near 0 across all redshift range, with $1\sigma$ scatter consistent with 0 for every bin. 

We define a catastrophic photometric redshift outlier to be where $|\Delta z|/(1+z_{H7}) > 0.1$. Inspecting the catastrophic outliers, we find some mass dependence, with photo-zs tending to underestimate the redshift for low-mass galaxies and overestimate for high-mass. This can be explained by the high-mass galaxies including a number of broad-line MgII emitters, which are rare and may not be well-represented in the template spectra used in the photo-z fitting, and by the possible degeneracy between low-z quiescent or dusty galaxies and high-z dropouts. For the low-mass catastrophic outliers, the photo-zs are often very near $z=0$, sometimes an indicator that the fit failed in an early step or perhaps the influence of any brightness-based assumptions, which place a lower probability on fits that imply an object is unusually bright or faint for its redshift. Finally, catastrophic outliers in the photo-zs are more common in those COSMOS and EGS galaxies outside of CANDELS, which may lack as comprehensive photometric coverage.

The overall fraction of catastrophic outliers is 5.5\% for the sample with good HALO7D redshifts, photometric redshifts, and stellar masses. Among the low-mass sample, this outlier rate is 7.5\%, while catastrophic outliers make up only 3.5\% of the high-mass group. These results show that while photometric redshift methods do not perform dramatically worse with low-mass galaxies, they do still produce higher rates of error which may be improved in the future with more comprehensive template libraries for dwarf galaxies or similar modifications to fitting procedures.
 
\section{Catalog Properties and The Dwarf Population} \label{sec:cat}

\begin{figure}
    \centering
    \includegraphics[width=0.5\textwidth]{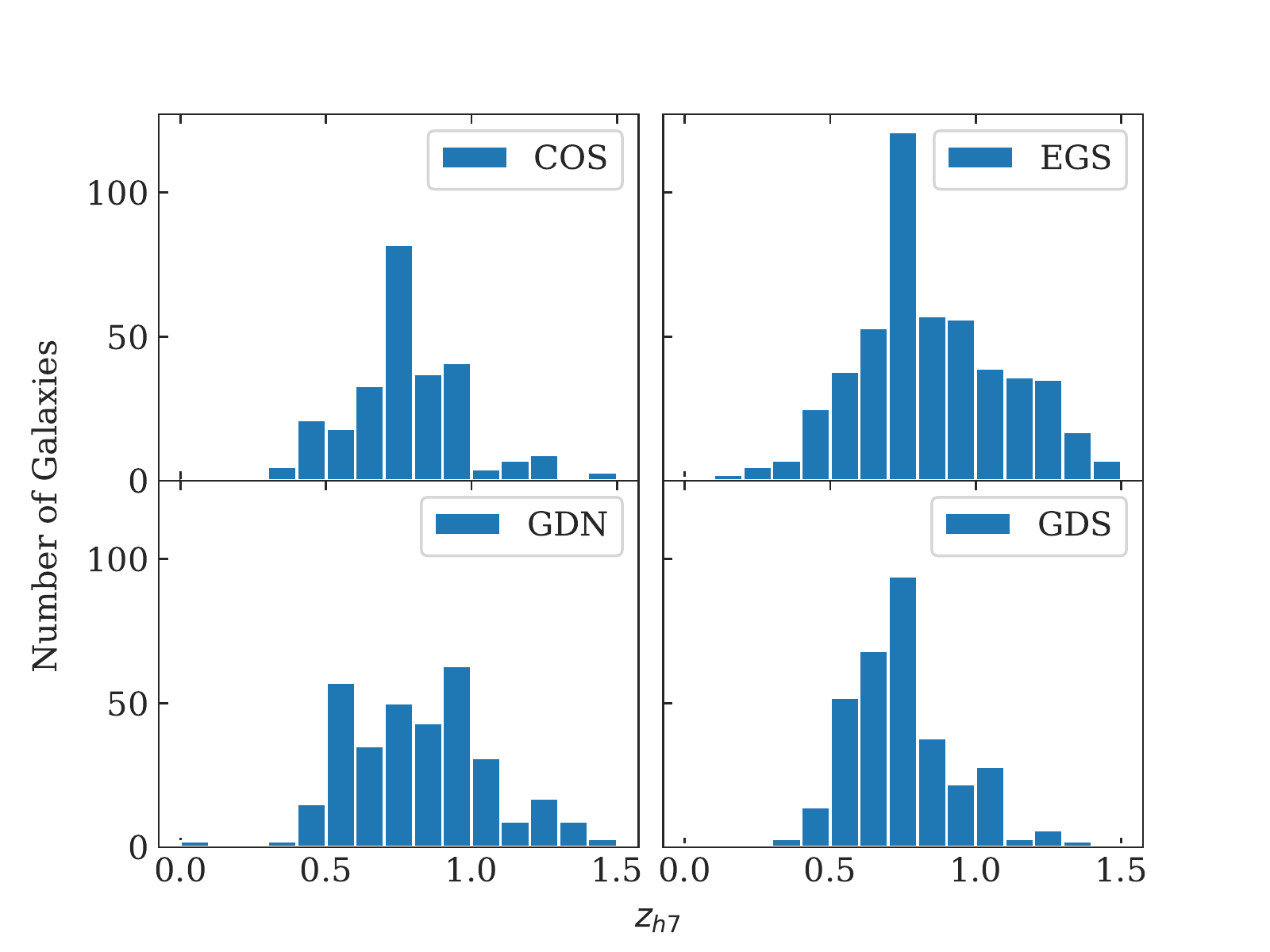}
    \caption{The distribution of $z_{h7}$ in each of the HALO7D fields.}
    \label{fig:z_hist}
\end{figure}

\begin{figure*}
    \centering
    \includegraphics[width=0.7\textwidth]{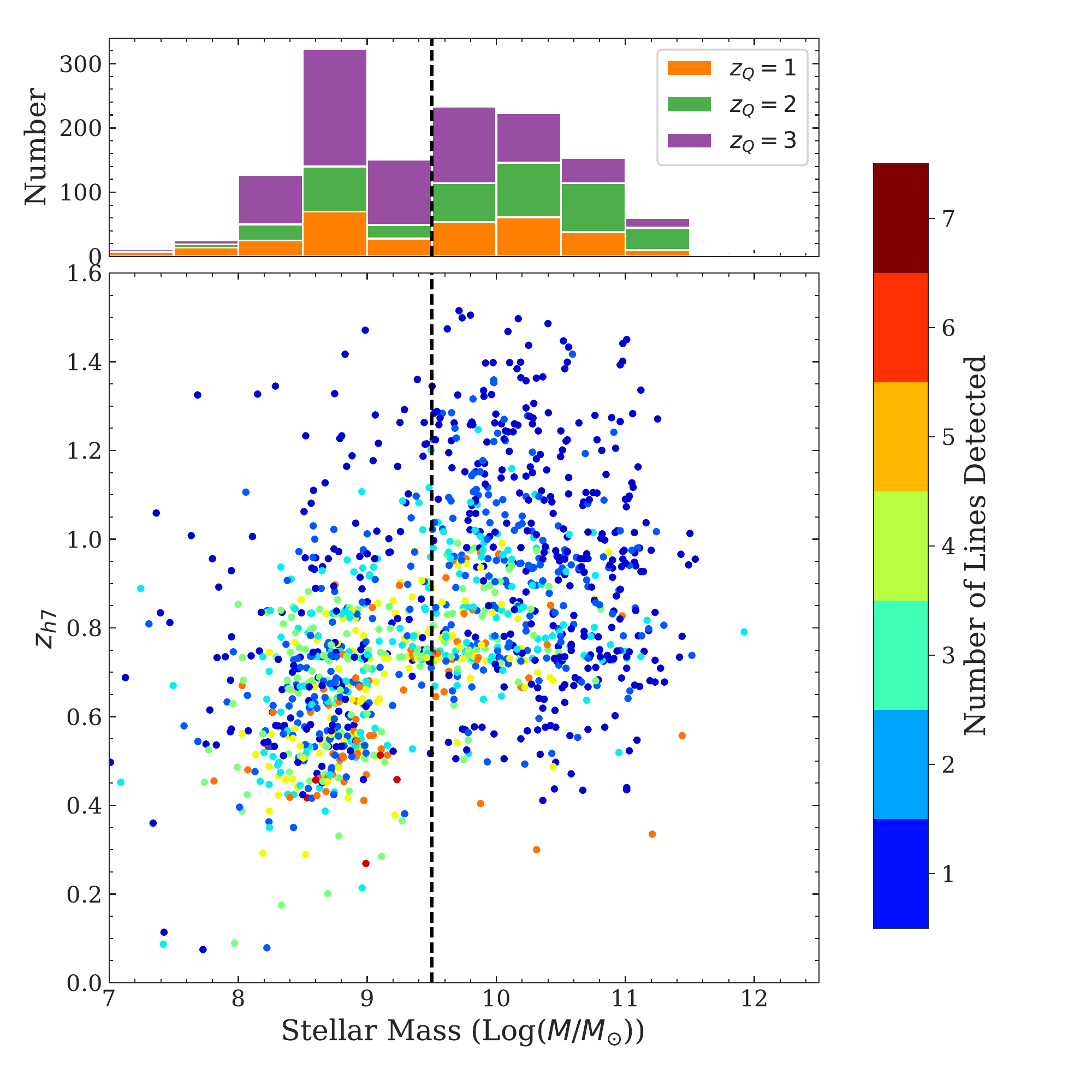}
    \caption{The redshift and stellar mass distribution of HALO7D redshift catalog. The black vertical dashed line demarcates the separation between the low-mass and high-mass samples at $\log(M_{\star}/M_{\odot}) = 9.5$. The points are colored by the number of significant features detected, counting the pair of Ca H and K absorption lines and the [O\textsc{ii}]3727,3729 doublet each as a single feature. The top panel gives a histogram of the mass distribution of the overall sample with good redshifts, separated by the redshift fit quality, $z_Q$.}
    \label{fig:mass_z}
\end{figure*}

\subsection{Redshift and Stellar Mass Distributions}

The redshift distribution of HALO7D galaxies is given in Figure \ref{fig:z_hist} for each of the four fields. Because the HALO7D targets were not uniformly selected, there is substantial variation in the redshifts measured in each field. The automated redshift grid produces good fits in the range $0 < z < 2$, with the vast majority of galaxies found in $0.4 < z < 1.5$, the redshifts at which [O\textsc{iii}]5007 moves into the blue end of the wavelength coverage and at which the [O\textsc{ii}]3727,3729 lines move out of the red end. This ignores the eight higher-redshift galaxies for which Mg II 2800 emission lines were identified visually, and are not included in this distribution. 

The redshift distributions tend to peak at $z \sim 0.8$, in part because at this redshift range all of [O\textsc{ii}], [O\textsc{iii}] and H$\beta$ can be detected, making secure redshift fits more probable, and in part because of target selection preferring such redshifts. The GDN field has a relatively broad distribution of redshifts compared to the other fields, and in comparison to the redshift distribution of DEIMOS-observed galaxies in the DEEP2 and DEEP3 surveys \citep{davis03,cooper11,newman13}. \citet{cooper11} shows a combined sample of 156 galaxies in GDN peaking notably at $z\sim0.5$, with only a handful of galaxies measured near $z\sim1$. The difference in the shape of the GDN distribution is this work is likely due to the overall increased size of the galaxy sample, combined with the survey depth enabling the measurement of a fainter galaxy population. The EGS field has a larger fraction of $z > 1$ galaxies compared to the others, likely a selection effect due to the relatively high fraction of higher-mass target galaxies in that field. 

Figure \ref{fig:mass_z} gives the overall redshift distribution of HALO7D galaxies as a function of their stellar mass in the lower panel. Galaxies are colored according to the number of significant features detected. The top panel gives a histogram of stellar masses, with the red dashed line demarcating the separation between the low-mass and high-mass samples at $\log(M_{\star}/M_{\odot}) = 9.5$. The low-mass sample is predominantly measured in the $0.4 < z < 1$ range, where detection of strong line emitters is most likely, while the high-mass population can be reliably measured at higher redshifts through detection of stronger Ca H and K absorption in older stellar populations. This distribution reflects the selection priorities for the survey, which included dwarf galaxies at a redshift where SF- and metallicity-tracing lines could be detected in the optical spectrum, high-z and high-mass SF galaxies that may exhibit strong galactic winds, and low-z and high-mass quiescent galaxies.

%, showing that the detection of multiple features is much more common at $0.4 < z < 1$.

\begin{figure*}
    \centering
    \includegraphics[width=\textwidth]{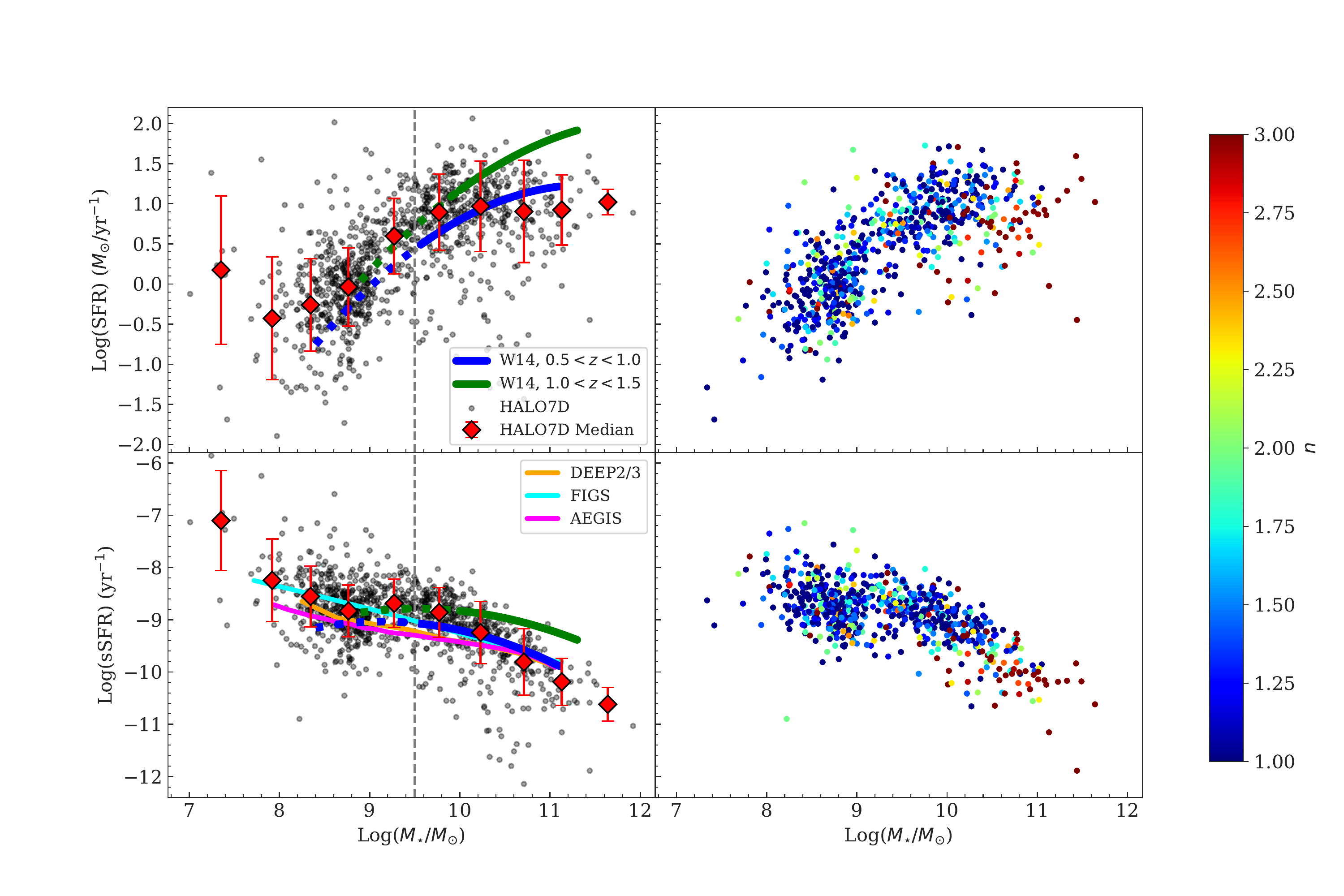}
    \caption{The SFR (top panel) and sSFR (bottom panel) of the sample as function of the stellar mass. In the left column, the vertical dashed line separates the dwarf sample from the massive galaxy sample, and individual HALO7D galaxies are given by black points, with dwarf galaxies filled and massive galaxies empty. Purple diamonds give the median SFR/sSFR for each mass bin, with error bars giving the standard deviation in the bin. We compare this with CANDELS/3D-HST SF-selected galaxies \citep{whitaker2014} at comparable redshifts ($0.5 < z < 1.0$ blue, $1.0 < z < 1.5$ green). These curves are each separated into parts, representing SFR measurements from stacking analysis (dotted) and individual SFRs (solid line). In the lower left panel, we include also relations from AEGIS \citep[magenta,][]{noeske07}, DEEP2/3 \citep[orange,][]{guo16b}, and FIGS \citep[cyan,][]{pharo20}. In the right column, the HALO7D points are colored by S{\'e}rsic index $n$ as determined in the CANDELS galaxy morphology catalogs from \citet{vanderwel2012}.}
    \label{fig:sfms}
\end{figure*}

\subsection{The Star-Forming Main Sequence and Galaxy Morphology}

\begin{figure*}
    \centering
    \begin{tabular}{cc}
        \includegraphics[width=0.43\textwidth]{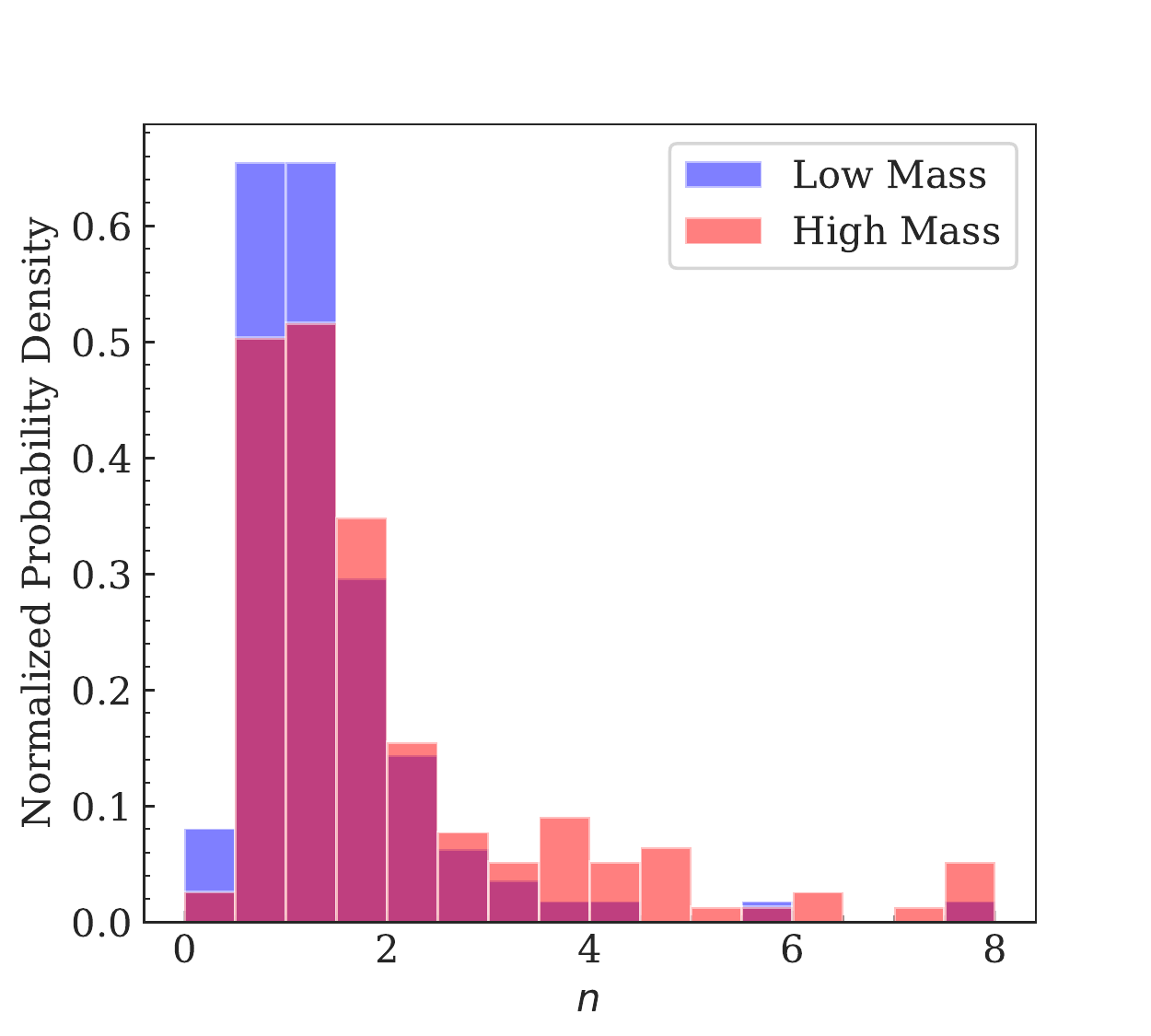} &  \includegraphics[width=0.6\textwidth]{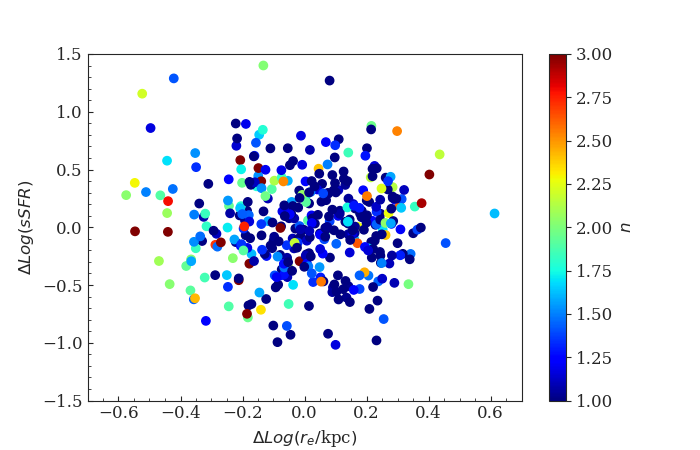}
    \end{tabular}
    
    \caption{\textit{Left:} The distribution of S{\'e}rsic indices from \citet{vanderwel2012} for the low-mass (blue) and high-mass HALO7D samples. \textit{Right:} The star formation rate residual $\Delta$log sSFR vs. the radius residual $\Delta$log $r_e$ for the dwarf galaxy sample, colored by S{\'e}rsic index $n$. The residuals are measured from the sSFR and size medians of the mass bins shown in Figure \ref{fig:sfms}. A Spearman rank correlation test finds a correlation coefficient consistent with 0 and no significant correlation measure. The color range of $n$ is restricted to best show possible variation among the predominantly low-$n$ dwarf sample.}
    \label{fig:sersic}
\end{figure*}

Hydrogen Balmer emission lines are commonly used to trace star formation (Kennicutt 1998). They are emitted through the recombination of ionized hydrogen, and so trace the ionizing radiation produced by recently formed, short-lived ($<10$ Myr) massive stars. Balmer emission lines in our catalog thus provide a way to analyze the star-forming properties of the dwarf galaxy population.

Balmer emission lines are also subject to absorption from the underlying stellar population, however, which will reduce the apparent flux if not corrected for. Stellar absorption can be measured by fitting a stellar continuum with absorption profiles, but this requires the spectra to observe the continuum with sufficient signal. For our dwarf-dominated galaxy sample, the stellar continua are typically quite low, so we instead adopt corrections from the literature for galaxies of similar mass and redshift. For galaxies with $Log(M_{\star}/M_{\odot}) > 9.5$, we adopt corrections of $EW_{abs}^{H\alpha}=3.4\AA$ and $EW_{abs}^{H\beta}=3.6\AA$ (Momcheva et al. 2018), and for galaxies with $Log(M_{\star}/M_{\odot}) \leq 9.5$, we use $EW_{abs}^{H\alpha}=EW_{abs}^{H\beta}=1\AA$ \citep{ly2015, guo16b}. We add these corrections to the observed Balmer EWs and adjust the measured line fluxes accordingly.

We then used the Balmer lines to measure dust extinction. For each galaxy with both H$\beta$ and $H\gamma$ significantly detected, we calculate the Balmer decrement and the E(B-V) extinction, then correct emission line fluxes using the E(B-V) measurement and the \citet{cardelli1989} extinction law. For galaxies without both detected Balmer lines, we use the median E(B-V) for galaxies in the same bin of stellar mass to estimate the extinction.

We calculate the SFR for HALO7D galaxies using the \citet{kennicutt98} formulation where $SFR = 7.9\times10^{-42}L_{H\alpha}$ and assuming Case B intrinsic flux ratios of $f_{Hn}/F_{H\alpha}$. SFR may also be calculated from calibrations of the [O\textsc{ii}] luminosity, but this is less precise as L(O[\textsc{ii}]) is sensitive to other parameters including the metallicity, and so is typically dependent on calibrations to Balmer emission \citep{kennicutt98}. Therefore, we use preferentially use Balmer lines to determine SFR, preferring the most intrinsically strong line detectable. Finally, we use the calibrations of \citet{kennicuttEvans2012} to adjust the SFR calculations from a Salpeter IMF to a Kroupa IMF, making the calculations more comparable to modern surveys, where the Kroupa or the similar Chabrier IMFs are most commonly used. %that more commonly use Kroupa or the similar Chabrier IMF.

Figure \ref{fig:sfms} shows the SFR and sSFR (SFR per stellar mass) plotted against the stellar mass, a relation commonly called the galaxy star-forming main sequence. The high-mass sample was not uniformly selected, and selection included targeting quiescent galaxies, which may explain the apparent flattening of the high-mass sample and why their SFRs often fall below those measured in the 3D-HST samples at comparable redshift \citep{whitaker2014}. At low mass, the median trend of the HALO7D sample is comparable to many previous surveys at similar redshifts \citep[AEGIS, DEEP2/3, FIGS;][]{noeske07, guo16b, pharo20}. Despite the depth of HALO7D enabling sSFR measurements down to $\sim-9.5$ in the dwarf galaxy sample, we do not observe as noticeable a flattening in the sSFR-$M_{\star}$ curve as in \citet{whitaker2014}, a UV-IR SFR study from 3D-HST. This is potentially a distinction between line-emission-selected SFRs in HALO7D, which typically probe very young starbursts and the UV+IR selected SFRs in the 3D-HST sample, which may be less prone to selecting recent bursts. This would also explain the consistency of our results with the other emission-line surveys. Furthermore, the \citet{whitaker2014} curves are separated into dotted and solid line regions, denoting the low-mass regime where their measurements are taken from stacked data, and the higher-mass regime where they made individual measurements. This shows that while HALO7D is able to probe individual measurements of star formation down to low levels consistent with the \citet{whitaker2014} curves, we still miss some low-SFR galaxies too faint for emission line detection that would still contribute to the stacks.

The right column of Figure \ref{fig:sfms} gives the same SFR and sSFR relations for HALO7D, with the points colored by the S{\'e}rsic index $n$ as fit by \citet{vanderwel2012}. There is an apparent relationship between $n$ and the stellar mass, which can be seen clearly in the left panel of Figure \ref{fig:sersic}, which gives the normalized probability density distributions of $n$ for both the dwarf and massive galaxy samples. This shows that the low-mass sample more strongly peaks at low $n \approx 1$, common to spiral and dwarf elliptical galaxies. The high-mass sample has a more substantial tail to high values of $n$, indicating a higher concentration of light in the center of galaxies representative of elliptical galaxies.

In order to check for a possible correlation between sSFR and galaxy size and morphology, we measure residuals from the Main Sequence and size-mass relations by subtracting individual sSFR and size measurements from the median sSFR and size from the matching mass bins shown in Figure \ref{fig:sfms}. The right panel of Figure \ref{fig:sersic} shows the star formation rate residual $\Delta$log sSFR vs. the radius residual $\Delta$log $r_e$ for the dwarf galaxy sample, colored by S{\'e}rsic index $n$. A Spearman rank correlation test finds a correlation coefficient $r$ consistent with 0 and no significant correlation between the two residuals, suggesting no clear relationship between star formation and galaxy size in dwarf galaxies at $0.5 < z < 1$. This corroborates a recent result in \citet{lin2020}, who similarly find no significant correlation in UV-IR SFRs and size in CANDELS and SDSS galaxies spanning $0 < z < 2.5$. We do find higher $n$ values associated with scatter in the size-mass relation, with median $n$ value of 2.0 and 1.8 in the two extreme $\Delta Log(r_e/$kpc$)$ bins, and only 1.2 and 1.0 in the middle bins.

\subsection{Color and Magnitude Properties}

\begin{figure}
    \centering
    \includegraphics[width=0.5\textwidth]{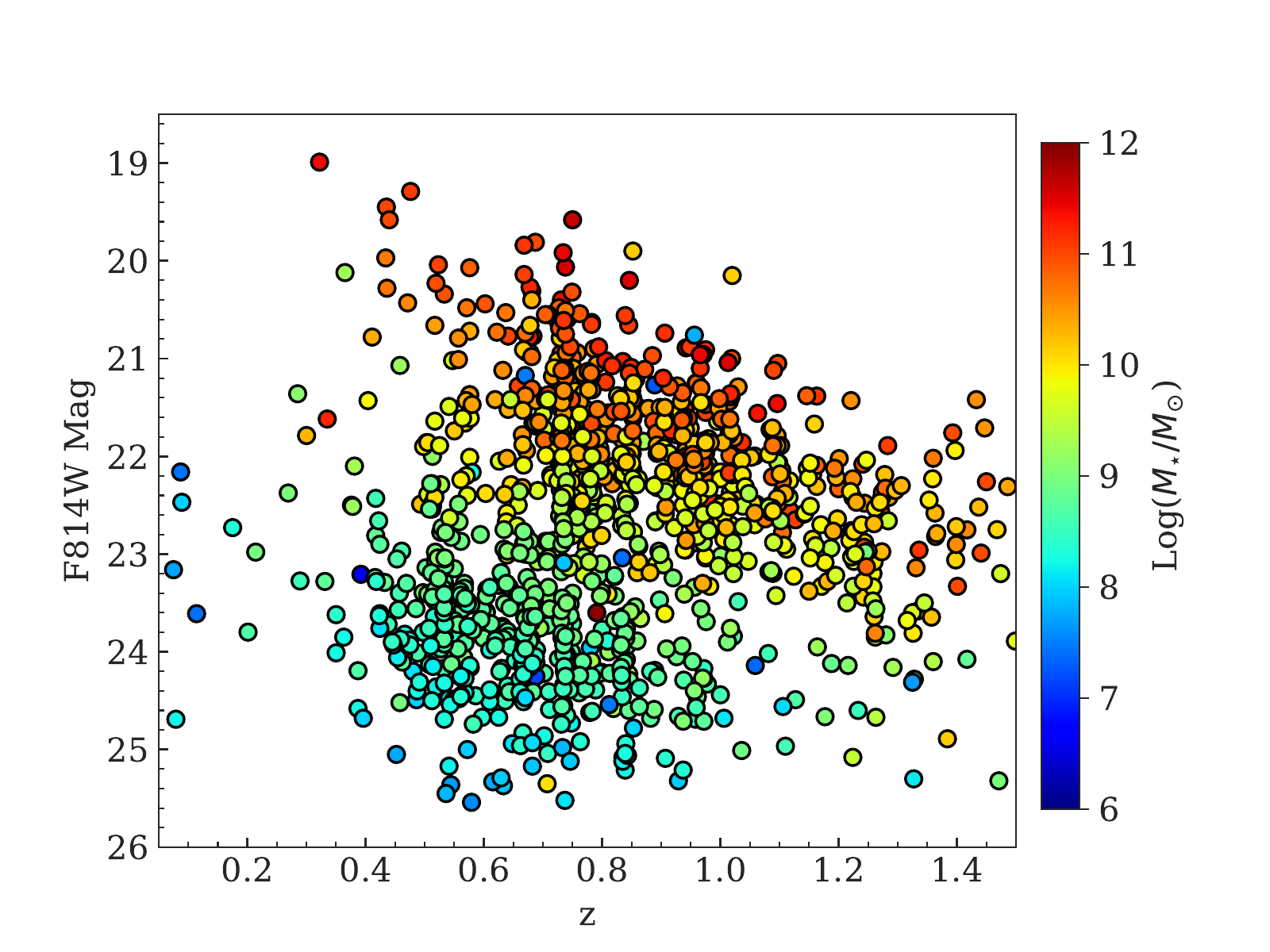}
    \caption{The F814W magnitude as a function of redshift for the HALO7D galaxies, colored by stellar mass.}
    \label{fig:colmag1}
\end{figure}

In this section, we explore the magnitude and color distributions of the HALO7D population for which we were able to measure redshifts from the HALO7D spectra. We compare the color-magnitude properties to the overall populations of galaxies in the CANDELS datasets at similar redshifts. Figure \ref{fig:colmag1} shows the distribution of F814W magnitudes for the catalog. F814W roughly corresponds to the redder continuum observed in the Keck/DEIMOS spectra. Plotted as a function of redshift and colored by stellar mass, this distribution demonstrates the three classes of galaxy targeted in the survey: dwarf galaxies at $z < 1$ ; massive, quiescent galaxies at $z < 1$ \citep{tacchella2021}; and massive galaxies that may host strong winds at $z > 1$ \citep{wang2021}. 

\begin{figure}
    \centering
    \includegraphics[width=0.5\textwidth]{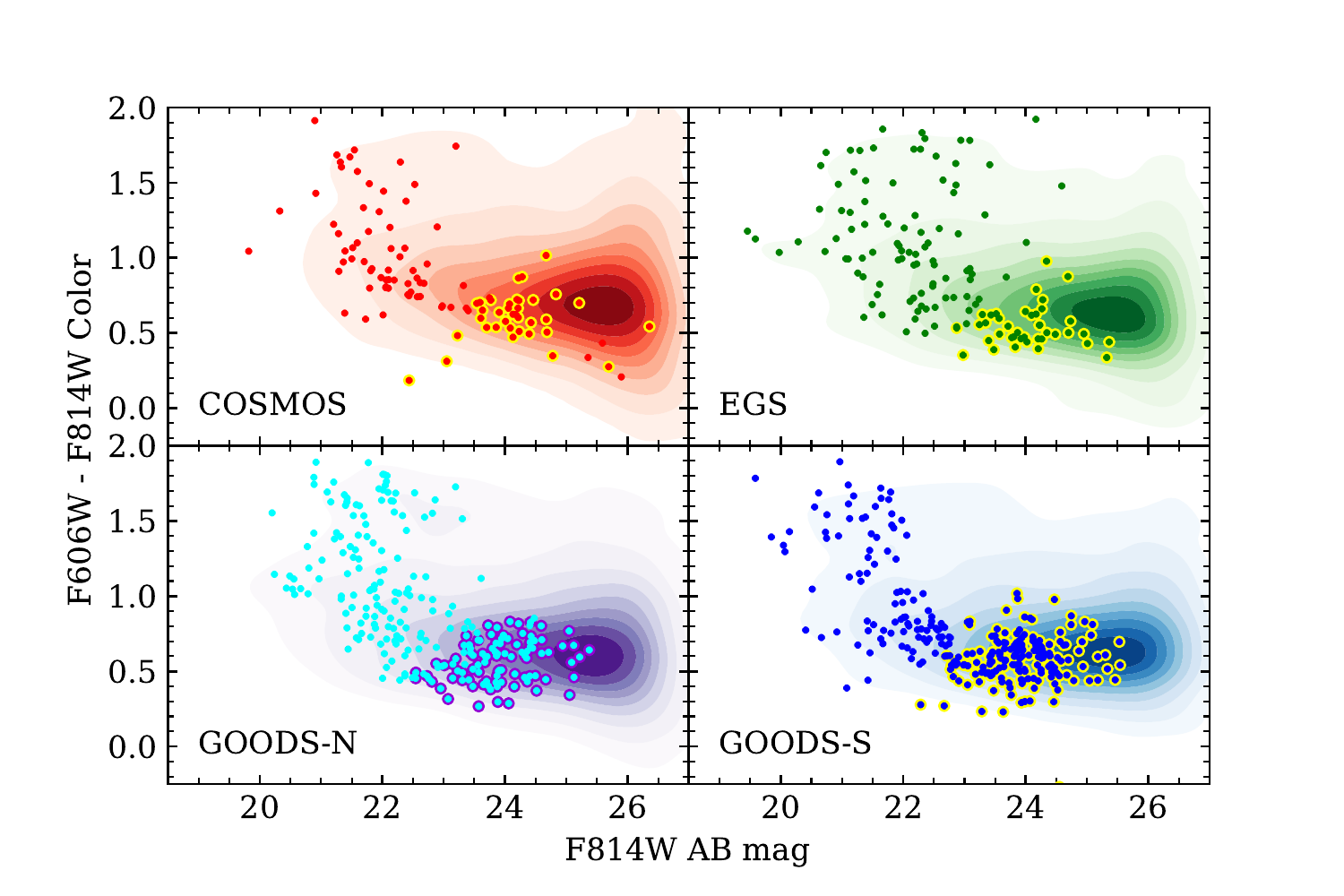}
    \caption{Observed-frame color-magnitude distribution of galaxies in each field with $0.4 < z < 1.0$, the redshift range where the most strong emission lines are likely to be detected in the DEIMOS optical spectra. HALO7D galaxies are colored points, with CANDELS galaxies shown in shaded contours. Dwarf galaxies in HALO7D are indicated by a colored border.}
    \label{fig:colmag2}
\end{figure}

In this paper, we largely focus on the dwarf galaxy population, defined to be those galaxies with $\log(M_{\star}/M_{\odot}) < 9.5$ and $0.4 < z < 1$, restricting the sample to low-mass galaxies in the redshift range where the most line identifications are possible (and, as can be seen in Figure \ref{fig:mass_z}, the region where the large majority of dwarf galaxies are identified). In Figure \ref{fig:colmag2}, we plot the F606W-F814W color against the F814W AB magnitude for this sample, broken down by field of observation. HALO7D galaxies are shown with colored points, and other CANDELS galaxies in the same redshift range are shown with gray points. At this redshift, this is approximately the rest-frame U-V color versus V magnitude. This helps to demonstrate the non-uniform selection of HALO7D galaxies, with the fainter (F814W mag $<$ 23) population primarily made up of dwarfs with a likely star-forming color. The color-magnitude distribution shows little differentiation from field to field.

Compared to the overall CANDELS galaxies, the HALO7D dwarf sample (indicated by colored outlines) occupies a narrow color range, with $0 <$ F606W-F814W $ < 1$ for virtually the whole sample. This is distinct from uniformly selected surveys such as DEEP2, where spectroscopic observations cover the full CANDELS color range \citep{wirth04}. This holds across the range of F814W magnitudes, which typically reach to 25 mag or fainter, which is $\sim1$ mag fainter than in \citet{wirth04}.

\begin{figure*}
\centering
\begin{tabular}{cc}
    \includegraphics[width=0.45\textwidth]{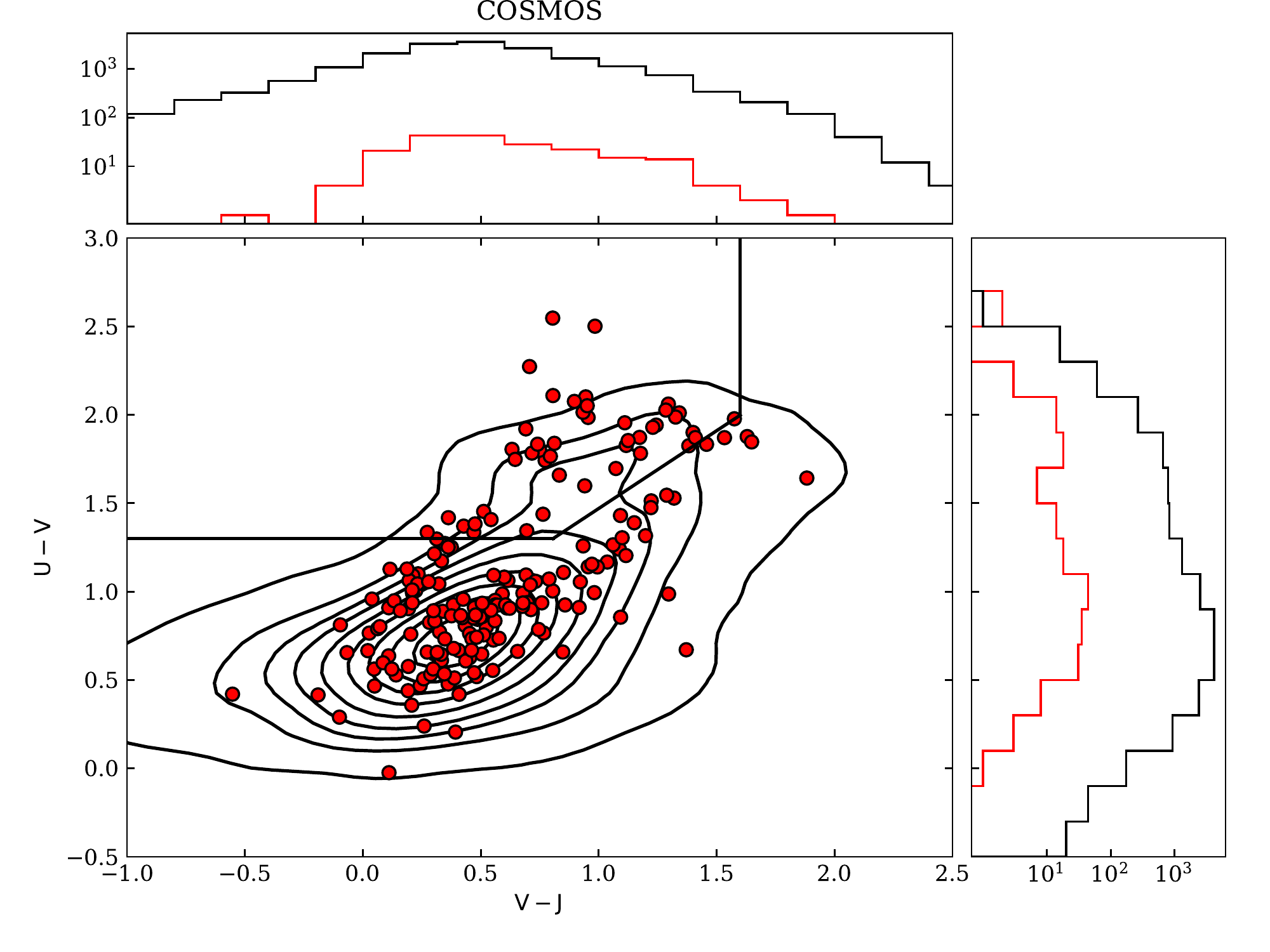} & \includegraphics[width=0.45\textwidth]{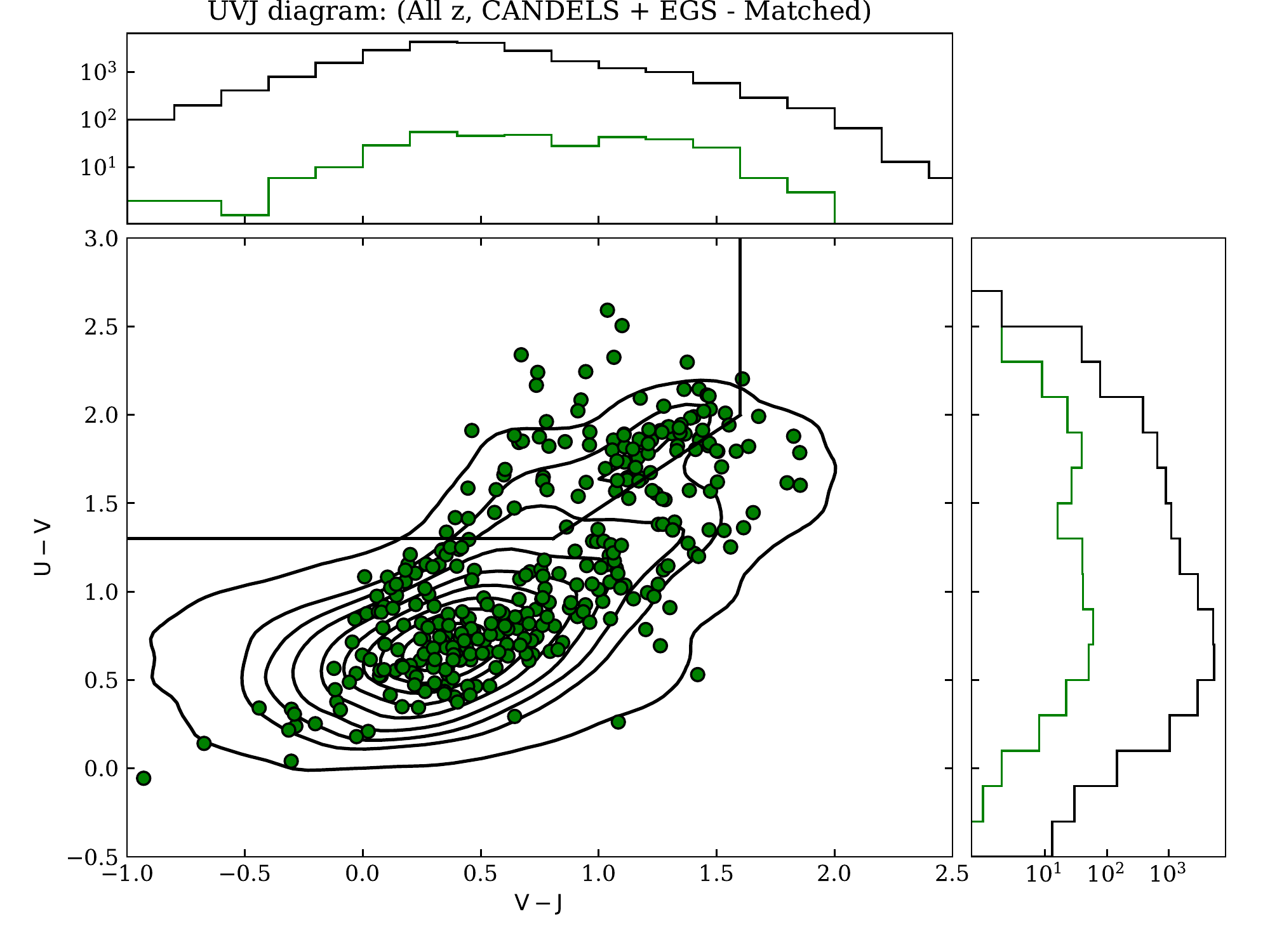} \\
    \includegraphics[width=0.45\textwidth]{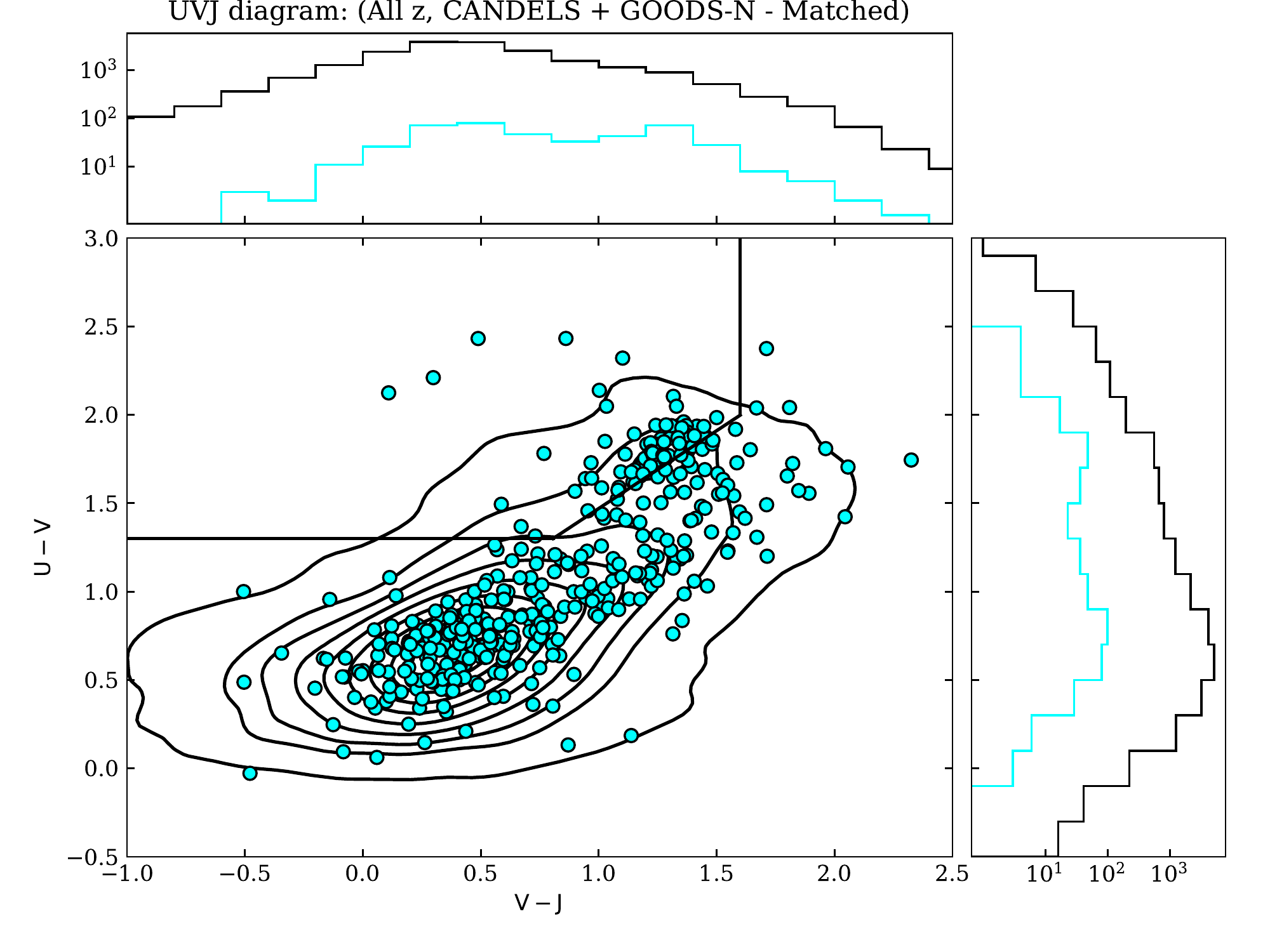} & \includegraphics[width=0.45\textwidth]{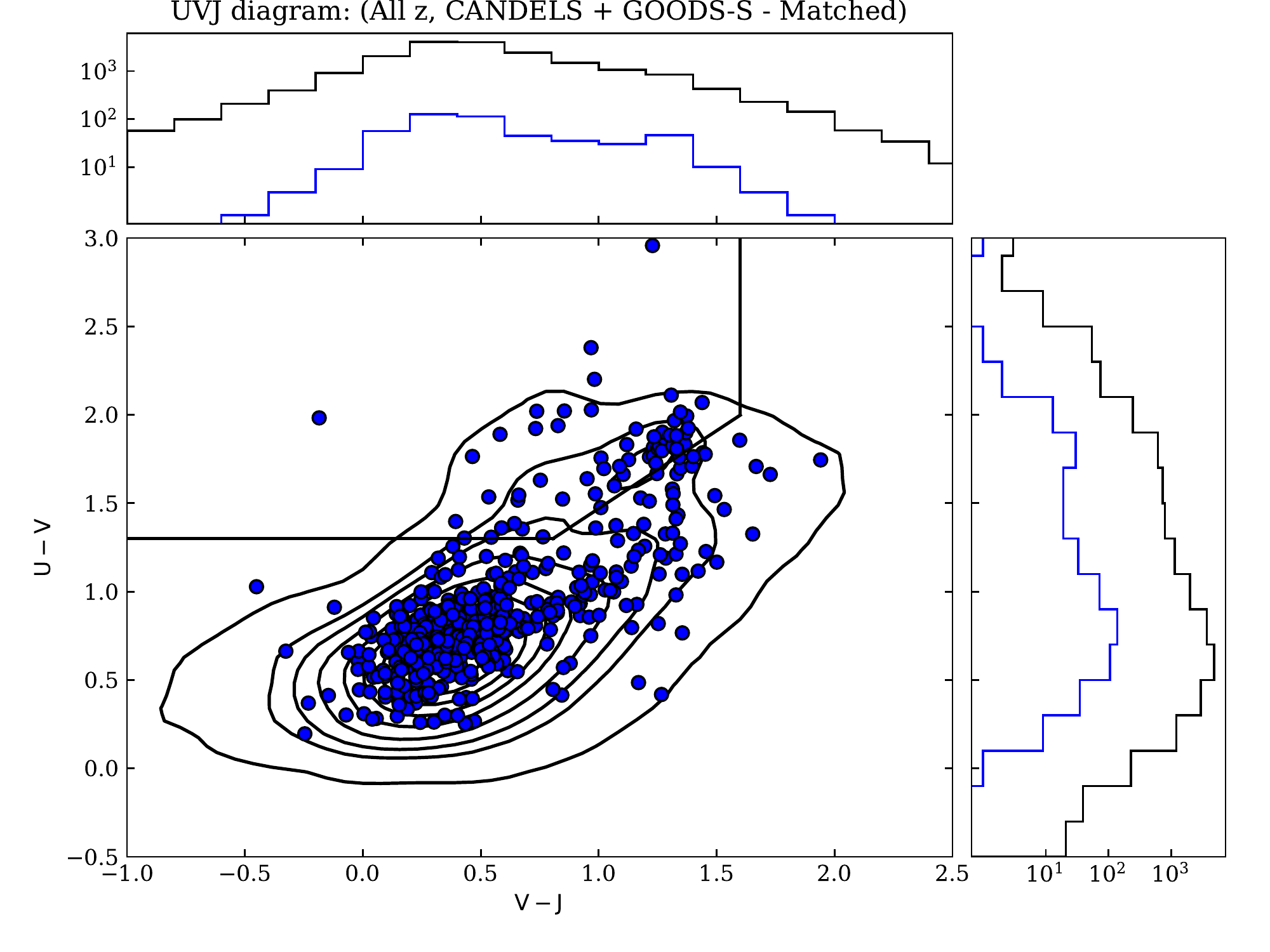}
\end{tabular}
\caption{U-V color versus V-J color distributions for galaxies in the four CANDELS fields observed in HALO7D (colored points) and in CANDELS (contours). The solid black lines separate the quiescent and star-forming galaxies, using the color selection criteria from \citet{williams2009} for $0.5 < z < 1.0$. The colors of each set of points are as in Figure \ref{fig:colmag2}.}
\label{fig:colcol}
\end{figure*}

\citet{kocevski17} calculates rest-frame U-V and V-J colors for the existing CANDELS catalogs, including many of the galaxies in the HALO7D sample. Figure \ref{fig:colcol} plots these rest-frame colors for the whole HALO7D and CANDELS samples in each field, with accompanying histograms giving the color distributions of each. With this parameter, the HALO7D sample distribution largely mirrors that of the overall CANDELS sample in most respects. One exception is a relative dearth of V-J $ < 0$ galaxies in HALO7D, especially visible in the COS field. This may be somewhat due to the filtering out of stellar objects from the HALO7D catalog, but may also speak to the lower success rate in measuring redshifts for low-mass galaxies in the COS field spectra. There is some appearance of a color bimodality in U-V, with a minimum at U-V$\approx 1.5$. This is similar to the U-B bimodality observed in \citet{wirth04} in DEEP2, \citet{hogg02} in SDSS, and \citet{bell04} in the COMBO-17 survey.

\begin{figure*}
    \centering
    \includegraphics[width=0.75\textwidth]{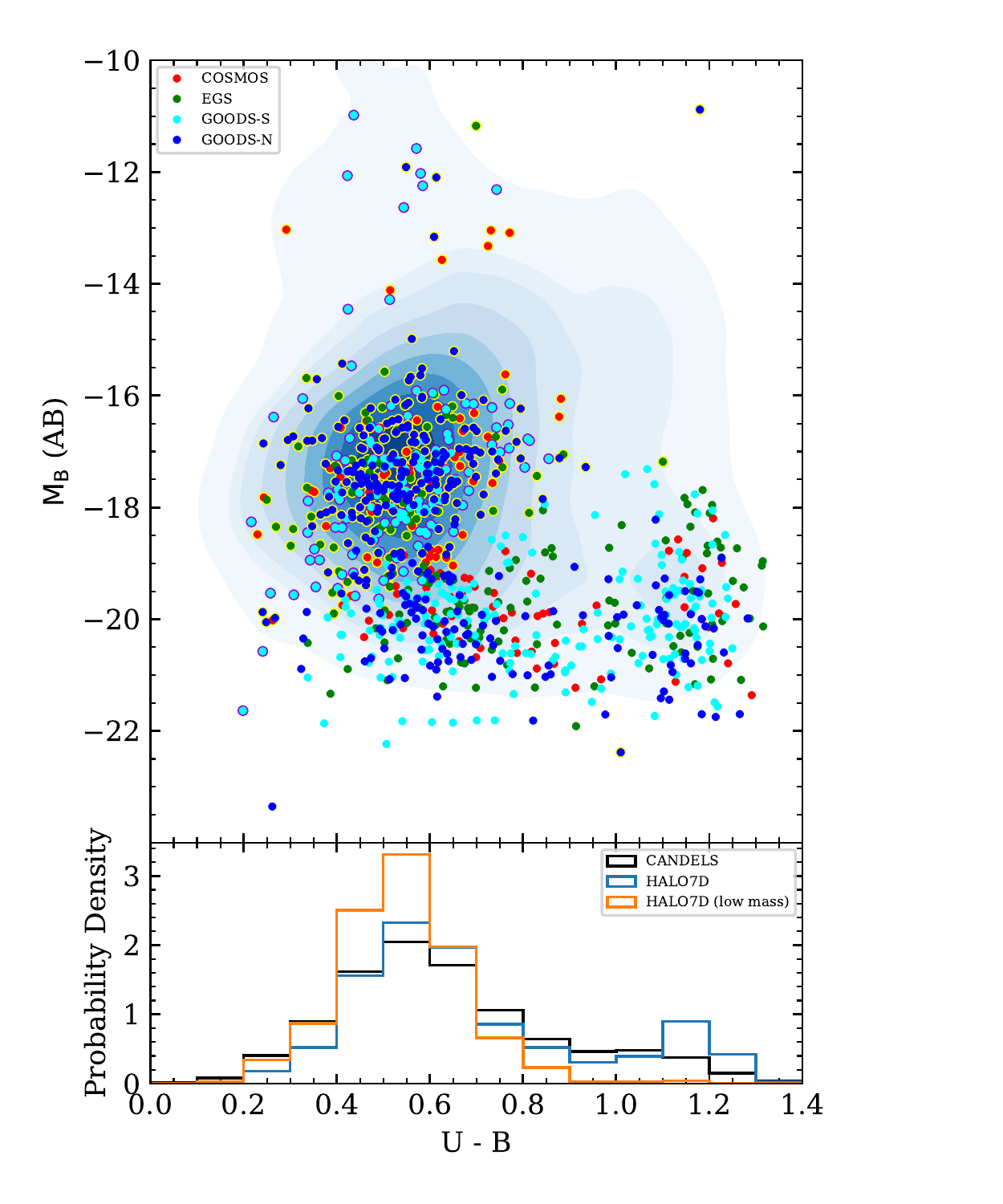}
    \caption{The rest-frame color-magnitude distribution of the HALO7D sample. HALO7D galaxies are colored according to their field, and the dwarf galaxies in the sample have colored outlines. The CANDELS galaxies from the same fields are shown as underlying shaded contours. The bottom panel shows the normalized probability density distribution of the rest-UB color in CANDELS, HALO7D, and the low-mass HALO7D sample. HALO7D galaxies overall have a comparable color distribution to CANDELS galaxies, and the dwarf sample predominantly features bluer UB color.}
    \label{fig:colmag3}
\end{figure*}

Using the rest-frame U and B band magnitudes for the HALO7D galaxies determined in the catalogs from \citet{kocevski17}, in Figure \ref{fig:colmag3} we give the rest-frame color-magnitude distribution for HALO7D. This distribution is plotted over the overall CANDELS galaxies in the HALO7D fields, and this demonstrates the non-uniformity of the HALO7D galaxy selection, especially in the low-mass sample. The U-B bimodality in the DEEP2 \citep{wirth04} rest-frame distribution can also be seen here in the overall HALO7D sample, and it somewhat corresponds to the low-mass and high-mass selections, with very few dwarf galaxies with $U-B > 0.9$. The galaxies with $U-B < 0.9$ are also shifted toward lower luminosities relative to similarly-colored galaxies in DEEP2, as is expected from the preferential targeting of dwarf galaxies.

\section{Summary and Conclusion} \label{sec:conc}

We have described the deep redshift survey of the COSMOS, EGS, GOODS-North, and GOODS-South CANDELS fields in HALO7D. With deep, up to 32-hour observations with the Keck/DEIMOS spectrograph, we analyzed optical spectra for $\sim2400$ galaxies. By identifying emission and absorption lines in the spectra, we fit redshifts to the spectra, obtaining 1440 redshift fits based on at least one spectral feature, two-thirds of which have more than one line detection. This also includes 646 galaxies with $\log(M_{\star}/M_{\odot} < 9.5)$, 454 of which did not previously have a published spectroscopic redshift. We present a catalog of these redshifts, including redshift quality flags and line detections.

We analyze the properties of this catalog, finding very close agreement with existing spectroscopic redshifts, with only 1\% of HALO7D redshifts differing from existing spec-zs by more than $0.01 \times (1+z_{spec})$. We obtained successful redshift fits for 75\% of massive galaxies across all fields, and for $\sim50\%$ of low-mass galaxies. We find that galaxies with $8.5 < \log(M_{\star}/M_{\odot}) < 9.5$ are fit successfully at rates comparable to the higher mass galaxies, with successful fits dropping substantially at $\log(M_{\star}/M_{\odot}) < 8.5$. This suggests we are able to measure galaxies with $8.5 < \log(M_{\star}/M_{\odot}) < 9.5$ at a level of completeness comparable to the high-mass sample.

We compare the redshift catalog to existing photometric redshifts, finding a median $|\Delta z| /(1+z_{h7})$ of 0.011 for the entire sample. We find that both the median error and the catastrophic outlier rate for photometric redshifts increase mildly with lower stellar mass, suggesting a need for slight modifications to redshift fitting methods to better account for the dwarf galaxy population, but that current photometric redshifts are still broadly viable for this population. Finally, we present and examine the redshift, mass, star-formation, morphology, and color properties of the non-uniformly selected HALO7D sample, noting any deviations in particular fields. We find no correlation between scatter in the star-forming main sequence and scatter in the size-mass relation for dwarf galaxies.

The redshift and emission line catalogs derived and presented here establish a dataset that will subsequently be used to measure gas-phase metallicities for the dwarf galaxy sample, probing the Mass-Metallicity Relation down to lower stellar mass at $z\sim0.7$ and investigating its relationship with star formation. This data may also be used in studies of the dwarf population's star formation properties, such as burstiness or links to galaxy morphology.

\acknowledgments

We would like to thank the anonymous referee for their many helpful comments. JP and YG would like to acknowledge support from NASA’s Astrophysics Data Analysis Program (ADAP) grant number 80NSSC20K0443. SAK and WW would like to acknowledge support from NASA ADAP grant number 80NSSC20K0760. We acknowledge support from NSF grant AST-1615730. This research made use of Astropy, a community-developed core Python package for Astronomy (Astropy Collaboration et al. 2013, 2018). We recognize and acknowledge the significant cultural role and reverence that the summit of Maunakea has always had within the indigenous Hawaiian community. We are most fortunate to have the opportunity to use observations conducted from this mountain.

\bibliography{refs}{}
\bibliographystyle{aasjournal}



\end{document}